\documentclass[a4paper,11pt]{article}
\usepackage{amsmath}
\usepackage{amsfonts}
\usepackage{amssymb}
\usepackage{graphicx,epsfig}
\usepackage{hyperref}
\usepackage{mathtools}
\numberwithin{equation}{section}
\usepackage[dvipsnames]{xcolor}
\usepackage{caption}
\usepackage{enumitem}
\usepackage{pifont}
\usepackage{array}
\usepackage{varwidth}
\usepackage{scalerel}
\usepackage{lscape}
\usepackage{afterpage}
\makeatletter
\renewcommand*\env@matrix[1][\arraystretch]{%
	\edef\arraystretch{#1}%
	\hskip -\arraycolsep
	\let\@ifnextchar\new@ifnextchar
	\array{*\c@MaxMatrixCols c}}
\makeatother 

\begin{document}
	\title{Bi-parametric $su(1,1)$ structure of the Heun class of equations and quasi-polynomial solutions 
		}
		\author{Priyasri Kar\\[0.2cm]
		{\it Department of Physical Sciences,}\\ {\it Indian Institute of Science Education and Research Kolkata, Mohanpur 741246, India}\\[0.2cm]
		{\tt pk12rs063@iiserkol.ac.in}
	}
	\maketitle
	\begin{abstract}
		A new bi-parametric $su(1,1)$ algebraization of the Heun class of equations is explored. This yields additional quasi-polynomial solutions of the form $\{z^{\alpha}P_N(z): \ \alpha \in \mathbb{C}, \ N \in \mathbb{N}_0\}$ to the General Heun equation and its confluent versions. Explicit conditions leading to these quasi-polynomials have been provided for the individual equations to allow direct use. For the Confluent and the Doubly-confluent Heun equations, specific parametric situations leading to (i) an infinite number of quasi-polynomials and (ii) non-algebraizability of the equation have been identified.
	\end{abstract}

	\section{Introduction}\label{sec_intro}
	
	Heun class of differential equations are found to appear in a varied range of physical problems~\cite{hortacsu_main,ronveauxbook} and have been studied extensively in the past few decades~\cite{ronveauxbook, slavyanov_lay, maier}. The General Heun equation (GHE) is a second order Fuchsian equation with four regular singularities. 
	The canonical form of this equation is given by
	\begin{equation}\label{ghe_canonical}
	\frac{d^2y(z)}{dz^2}+\left(\frac{\gamma}{z}+\frac{\delta}{z-1}+\frac{\varepsilon}{z-a}\right)\frac{dy(z)}{dz}+\frac{\alpha\beta
		z-q}{z(z-1)(z-a)}y(z)=0,
	\end{equation}
	with regular singularities at $z$ = $0, \ 1, \ a(\ne 0,1) \mbox{ and} \ \infty$. The confluent versions of this equation are obtained by the coalescence of two or more regular singular points in different possible ways. For the differential equations appearing in physics, the global solutions, valid throughout the entire finite region of the complex plane (with proper branch cuts to avoid the singularities), are of great importance in theoretical studies. The most useful way to obtain global solutions is factorization. The hypergeometric equation (second order Fuchsian equation with three regular singularities at $z$ = $0, \ 1 \mbox{ and} \ \infty$), of which the general Heun equation is the natural extension, was factorized by Schr\"{o}dinger~\cite{schroedinger} long time back and also all the second order equations equivalent to this equation are known to be factorizable~\cite{infeld_hull}. However, the factorization of the GHE or its confluent versions for general parameter values remains elusive till date (although some attempts reveal that they are factorizable when subjected to certain parametric conditions~\cite{heun_factorization, heun_cnflnts_factorization}). The next type of global solutions that one can look for, are the terminated Frobenius series solutions, which are quasi-polynomials (a superset of polynomials) of the form $\{z^{\alpha}P_N(z): \ \alpha \in \mathbb{C}, \ N \in \mathbb{N}_0\}$.
	The polynomial solutions of the Heun class of equations have been the subject of many investigations~\cite{turbiner_big,ciftci_hall_saad_dogu,gurappa_pkp_jpa,shapiro1,shapiro2,fiziev_che}. These equations are known to be quasi-exactly solvable~\cite{turbiner_big,turbiner,shif_tur,ush_book,lopez_kamran,proy,ishkhanyan_heun_schrodinger}, for a part of the eigenspectrum, consisting of polynomials, can be analytically tracked by exploring the algebraic properties of the equations. These equations can be cast in terms of the following set of $su(1,1)$ generators of \emph{degrees}\footnote{The degree $d$, of an operator $O_d$, is defined as the change in the power of a monomial, when acted upon by it, i.e., $O_d \ z^p \propto z^{p+d}$. } $\pm 1$ and $0$, 
	\begin{equation}\label{jpjzjm}
	\mathcal{J}^+= z^2\dfrac{d}{dz}-2j z, \quad
	\mathcal{J}^0= z\dfrac{d}{dz}-j \quad \mbox{and} \quad 
	\mathcal{J}^-= \dfrac{d}{dz},
	\end{equation}
	parameterized by a single parameter $j$, proposed for the first time by Sophus Lie. A detailed analysis is available in Ref.~\cite{turbiner_big}. Recently, a different bi-parametric set of $su(1,1)$ generators of degrees $\pm 1/2$ and $0$ has been used to cast a subset of the General Heun operator, with elementary singularities at the points zero and infinity~\cite{degree_half}. This has yielded some additional quasi-polynomial solutions (of the form $z^{\alpha}P_N(z)$, $\alpha \neq 0$) to the GHE, which are known as the \emph{Heun polynomials}~\cite{ronveauxbook}.
	It may seem tempting to ask whether there exists a more general algebraic structure, that can provide additional quasi-polynomial solutions to the GHE with general singularity structure and to the confluent versions of the GHE as well.
	
	The present work provides a positive answer to the above question. The following bi-parametric set of $su(1,1)$ generators
	\begin{subequations}\label{newgenerators}
		\begin{align}
		J^+=& z^2\dfrac{d}{dz}-2\sigma z, \label{newjp} \\
		J^0=& z\dfrac{d}{dz}-(\sigma+\tau) \label{newj0} \\ \mbox{and} \quad 
		J^-=& \dfrac{d}{dz}-\dfrac{2\tau}{z}, \label{newjm}
		\end{align}
	\end{subequations}
	of degrees $+1$, $0$ and $-1$ respectively, is proposed for algebraization. It may be noted that when one of the two parameters $(\tau)$ is set to zero, the above set of generators reduces to the generators~(\ref{jpjzjm}). The generators~(\ref{newgenerators}) satisfy the $su(1,1)$ commutation relations
	\begin{equation}\label{su11}
	\left[J^0,J^{\pm}\right]=\pm J^{\pm}, \qquad  \left[J^+,J^-\right]=-2J^0
	\end{equation}
	in the monomial space and the quadratic Casimir operator is given by
	\begin{equation}\label{construct_casimir}
	C_2=\dfrac{1}{2}\left[J^+J^-+J^-J^+\right]-J^0J^0=-\left(\sigma-\tau\right)\left(\sigma-\tau+1\right)=-j(j+1),
	\end{equation}
	where $j=(\sigma-\tau)$. Evidently the Casimir is a constant for specific $\sigma$ and $\tau$. In the subsequent sections it will be observed that-- (i) the General Heun equation and all of its confluent versions can be cast in terms of these newly proposed generators (modulo very restrictive parametric situations for the Confluent (CHE) and the Doubly-confluent (DHE) Heun equations) and (ii) for the General, Confluent, Bi-confluent and Doubly-confluent Heun equations, the new set of generators opens up possibility of obtaining some quasi-polynomial solutions that are not directly tractable using the generators~(\ref{jpjzjm}). Parametric relations are identified, once again for the CHE and the DHE, for which these equations are \emph{exactly solvable} (defined in section~\ref{subsec_heun_polynomials}). One must be convinced that the coverage provided by the new set of generators, both in terms of the number of algebraizable equations and the number of solutions obtained using the algebra, must incorporate the coverage of the generators~(\ref{jpjzjm}) as a subset, since the new set itself incorporates the old one as a special case which is obtained with the choice $\tau=0$. In fact, it will be found to be a proper subset since (i) some of the obtained solutions will be found to be associated with non-zero values of $\tau$ and (ii) it will be found that there exists Heun class equation (the DHE), the very algebraization of which (at least directly without any transformation) demands non-zero $\tau$.
	
	The organization of the paper is as follows: In section~\ref{sec_our_gens_&_heun_class} the Heun class of equations is cast in terms of the generators~(\ref{newgenerators}) and the algebraizability conditions are identified followed by the conditions of quasi-exact and exact solvability of the Heun class of equations in terms of quasi-polynomial solutions. In section~\ref{sec_eqwise_analysis}, the general findings of section~~\ref{sec_our_gens_&_heun_class} are applied to the individual equations of the Heun class and the explicit conditions for obtaining the quasi-polynomials are identified in each case. Finally, in the concluding section~\ref{sec_conclusion} the results are summarized and some physics applications of the present work are indicated. The present method involves exploring the $su(1,1)$ structure of the Heun class of equations and exploiting the $su(1,1)$ representation theory to obtain solutions. The appendix~\ref{appendix_rep_theory} gives a necessary account of the $su(1,1)$ representation theory. The appendix~\ref{appendix_imp_connections} provides important insights into the connections of the solutions obtained using present method to those obtained using the generators~(\ref{jpjzjm}) and to the standard Frobenius series solutions. The matrix forms of the GHE, CHE and BHE when they admit $(N+1)$ linearly independent polynomial solutions, as have been discussed in Ref.~\cite{ciftci_hall_saad_dogu}, have been provided in appendix~\ref{appendix_poly_conds} for the sake of completeness of the present work.
	
	\section{The Heun class of operators and the new $su(1,1)$ generators}\label{sec_our_gens_&_heun_class}
	
	The Heun class of equations consists of the General Heun equation (GHE) and its confluent versions, viz, the Confluent (CHE), the Bi-confluent (BHE), the Doubly confluent (DHE) and the Tri-confluent (THE) Heun equations. The General Heun equation is a second order Fuchsian equation with four regular singularities, whereas, the confluent versions are obtained by the coalescence of two or more regular singularities of the GHE. It may be noted that several other widely known equations making frequent appearance in physics (e.g. the Lam\'{e} equation, the Mathieu equation, the Legendre equation, the Jacobi equation, the General Spheroidal equation, the Coulomb Spheroidal equation etc.) are connected to the Heun class of equations either by specific choice of parameters or by simple variable transformations or both. Hence, the study of the properties and solutions of the Heun class actually helps to study all these equations analytically.
		
	The canonical form of the Heun class of equations is given by~\cite{ronveauxbook}
	\begin{equation}\label{heun_as_poly}
	\mathcal{H}y(z)=\left[P_3(z)\dfrac{d^2}{dz^2}+P_2(z)\dfrac{d}{dz}+P_1(z)\right]y(z)=0
	\end{equation}
	where
	\begin{equation}\label{the_P's}
	P_3=a_{0}z^3+a_{1}z^2+a_{2}z+a_3, \quad P_2=a_{4}z^2+a_{5}z+a_{6} \quad \mbox{and} \quad P_1=a_{7}z+a_{8}+\dfrac{a_9}{z},
	\end{equation}
	the $a_{i}$'s $\{i = 0,\ldots,9\} $ being arbitrary constants among which, $-a_8$ is clearly the eigen-parameter. The Heun operator $\mathcal{H}$ can be further rewritten as \cite{Arunesh}, 
	\begin{equation}\label{heun_as_cubic_sl2}
	\mathcal{H}=O^+ + O^0 + O^-+O^{--}
	\end{equation}
	where,
	\begin{subequations}\label{op_o0_om}
		\begin{align}
		O&^+= a_{0}z^3\frac{d^2}{dz^2}+a_{4}z^2\frac{d}{dz}+a_{7}z,\label{op}\\ 
		O&^0= a_{1}z^2\frac{d^2}{dz^2}+a_{5}z\dfrac{d}{dz}+a_{8},\label{o0}\\ 
		O&^-= a_{2}z\frac{d^2}{dz^2}+a_{6}\frac{d}{dz}+\dfrac{a_9}{z}\label{om}\\
		\mbox{and} \quad O&^{--}=a_3\dfrac{d^2}{dz^2}.\label{omm}
		\end{align}
	\end{subequations}
	The operators $O^i \{i=+,0,-,--\}$ are of degrees $+1$, $0$, $-1$ and $-2$, respectively. The following structure is proposed to cast the Heun operator $\mathcal{H}$ in terms of the $su(1,1)$ generators~(\ref{newgenerators}):
	\begin{equation}\label{heun_as_sl2}
	\mathcal{H}=\sum_{\substack{k,l=+,0,-{}\\k \geq l}}c_{kl}J^kJ^l +\sum_{k=+,0,-}c_kJ^k +c,
	\end{equation}
	where $c_{kl}, c_k$ and $c$ are arbitrary constants. Observation of Eqs.~(\ref{op_o0_om}) immediately tells that the coefficient $c_{++}$ is zero for the Heun class of equations. The coefficient $c_{00}$ is same as $c_{+-}$ apart from constant which can be absorbed in $c$, so one can effectively begin with $c_{00}=0$. The coefficient $c_{--}$ may have non-zero value but for just a single equation of Heun class (the THE), as will be seen in section~\ref{sec_eqwise_analysis}. If the concerned Heun class operator has the proposed $su(1,1)$ structure~(\ref{heun_as_sl2}) (or, is \emph{algebraizable} in terms of generators~(\ref{newgenerators})), then it should be possible to express the equation parameters $a_i$ (which are known from the physics problem in question) in terms of the constants $c_{kl},\ c_k$, $c$ and the two parameters $\sigma$ and $\tau$ associated with the generators~(\ref{newgenerators}). The following discussion focuses on two main features of the Heun class of equations: \textbf{algebraizability} and \textbf{(quasi-)exact solvability}.
	
	\subsection{The $su(1,1)$ Algebraizability of the Heun class of equations:}\label{subsec_algebraizability}
	
	The different parts of the equation given by Eqs.~(\ref{op}) to (\ref{omm}) are analysed separately for the algebraizability condition (if any) arising out of them.
	\begin{itemize}[label=\ding{228}]
		\item From the $O^+$ part (Eq.~(\ref{op})) one obtains,
		\begin{subequations}\label{a0a3a6}
			\begin{align}
			a_0=& c_{+0}, \label{a0} \\
			a_4=& c_++c_{+0}\left(1-3\sigma-\tau\right) \label{a3} \\ \mbox{and} \quad 
			a_7=& -2\sigma \left[c_+-c_{+0}\left(\sigma+\tau\right)\right] \label{a6}\\
			=& -2\sigma\left[a_4-a_0(1-2\sigma)\right]. \label{a6_alternative}
			\end{align}
		\end{subequations}

		Clearly, the self-consistency condition given by Eq.~(\ref{a6_alternative}) must be satisfied for algebraizability and it is satisfied \emph{except} for the very restrictive parametric situation
		\begin{equation}\label{non_alg_frm_op}
		a_0=a_4=0, \ a_7 \neq 0,
		\end{equation}
		for which no value of $\sigma$ will lead to non-zero $a_7$. Evidently, the non-algebraizability due to this particular parametric combination includes the case $\tau=0$ (Eq.~(\ref{a6_alternative}) is independent of $\tau$ anyway), which corresponds to the generators~(\ref{jpjzjm}). So it is found that there \emph{can be} parametric situations for which an equation of Heun class may \emph{not} have $sl(2,R)$ structure. The non-algebraizability, however, may occur for only two equations of the Heun class (the CHE and the DHE), since for the others at least one of $a_0$ and $a_4$ is necessarily non-zero (section~\ref{sec_eqwise_analysis}).
		
		For all other cases when Eq.~(\ref{non_alg_frm_op}) is not satisfied, $O^+$ is algebraizable. The type of singularity at $z=\infty$ turns out to be important for the determination of the parameter $\sigma$. The dependence is discussed below case by case:
	
		\textbf{$z=\infty$ is a regular singular point:}
		
		In this case one has $a_0 \neq 0$, which yields
		\begin{subequations}\label{sigma&cp}
			\begin{align}
			\sigma=&\dfrac{\left(a_0-a_4\right)\pm \sqrt{\left(a_0-a_4\right)^2-4a_0a_7}}{4a_0}, \label{ghe_sigma} \\ 
			\mbox{and} \quad c_+=&\dfrac{1}{4}\left[a_4+a_0\left(4\tau-1\right)\pm 3\sqrt{\left(a_0-a_4\right)^2-4a_0a_7}\right].\label{ghe_cp}
			\end{align}
		\end{subequations}
	
		\textbf{$z=\infty$ is an irregular singular point:}
		
		This corresponds to $a_0=0$. This may give rise to two different subcases:
		\begin{itemize}
			\item[$\bullet$] \textbf{$\sigma$ has a specific value:} This occurs when $a_4 \neq 0$. One obtains
			\begin{equation}\label{sigma_cnflnt}
			\sigma=-\dfrac{a_7}{2a_4} \quad \mbox{and} \quad c_+=a_4.
			\end{equation}
			\item[$\bullet$] \textbf{$\sigma$ is a free parameter:} This situation arises when $a_4=a_7=0$ in addition to $a_0=0$, i.e., the $O^+$ part is absent. In this case one has $c_{+0}, \ c_+=0$ and any value of $\sigma$ can be used for algebraization.
		\end{itemize}
		It may be noted that the parameter $\sigma$, associated with the generator $J^+$, is related to the exponent(s) of the local Frobenius series about $z=\infty$, the actual values of the exponents being given by $-2\sigma$ (appendix~\ref{appendix_imp_connections}). There are two of them (Eq.~(\ref{ghe_sigma})) when $z=\infty$ is a regular singular point and one of them (Eq.~(\ref{sigma_cnflnt})) when $z=\infty$ is an irregular singular point and $O^+$ is not absent. When $O^+$ is absent, the singularity at $z=\infty$ is regular, however, both the exponents of this regular singularity, being related to the eigen-value $-a_8$~(appendix~\ref{appendix_imp_connections}), are free parameters.
		
		\item Comparing the $O^-$ part (Eq.~(\ref{om})) with Eq.~(\ref{heun_as_sl2}) one obtains,
		\begin{subequations}\label{a2a5}
			\begin{align}
			a_2=& c_{0-}, \label{a2} \\
			a_6=& c_- -c_{0-}\left(\sigma+3\tau\right) \label{a5} \\ \mbox{and} \quad
			a_9=& -2\tau\left[c_--c_{0-}\left(1+\sigma+\tau\right)\right] \label{tau_condn}\\
			=& -2\tau\left[a_6+a_2(2\tau-1)\right]. \label{a9_alternative}
			\end{align}
		\end{subequations}
		
		Once again, the self-consistency condition~(\ref{a9_alternative}) must be satisfied for algebraizability. Here too non-algebraizability occurs for very restrictive parametric combination given by
		\begin{equation}\label{non_alg_frm_om}
		a_2=a_6=0, \ a_9 \neq 0,
		\end{equation}
		for which no value of $\tau$ can lead to non-zero $a_9$. This is true for any value of $\tau$, obviously including the case $\tau=0$ (corresponding to the generators~(\ref{jpjzjm})), vouching again for the fact that Heun class of equations does not necessarily have $sl(2,R)$ structure. However, this non-algebraizability may occur only for DHE, since for the others $a_9=0$ (section~\ref{sec_eqwise_analysis}).
		
		For all other cases when Eq.~(\ref{non_alg_frm_om}) is not satisfied, $O^-$ is algebraizable. The determination of the parameter $\tau$ turns out to be associated with the type of singularity at $z=0$. This is discussed below case by case.
	
		\textbf{$z=0$ is a regular singular point:} 
		
		This case corresponds to $a_2 \neq 0$ (and $a_3=0$) which leads to
		\begin{subequations}\label{tau&cm}
			\begin{align}
			\tau=&\dfrac{a_2-a_6 \pm \sqrt{(a_6-a_2)^2-4a_2a_9}}{4 a_2} \label{tau_regular} \\
			\mbox{and} \quad
			c_-=&\dfrac{1}{4} \left(a_6+a_2(3+4\sigma) \pm 3 \sqrt{a_6-a_2)^2-4a_2a_9}\right). \label{cm_regular}
			\end{align}
		\end{subequations}
		Eq.~(\ref{tau_regular} tells that at least one of the two $\tau$ value will in general be non-zero, which distinguishes the present set of $su(1,1)$ generators given by Eqs.~(\ref{newgenerators}) from the generators~(\ref{jpjzjm}). This is an extremely important finding, the impact of which is discussed in section~\ref{subsec_heun_polynomials}.
	
		\textbf{$z=0$ is an irregular singular point:}
		
		In terms of parameters this means $a_2=0$ (and $a_3=0$). Two subcases may arise just like the $O^+$ part:
		\begin{itemize}
			\item[$\bullet$] \textbf{$\tau$ has a specific value:} This occurs for $\ a_6 \neq 0$. One obtains
			\begin{equation}\label{the_thing_dhe}
			\tau=-\dfrac{a_9}{2a_6} \quad \mbox{and} \quad c_-=a_6.
			\end{equation}
			When additionally one has $a_9 \neq 0$, Eq.~(\ref{the_thing_dhe}) gives non-zero value for $\tau$ which can be attributed to the use of the generators~(\ref{newgenerators}). It may be noted that in this case, when $\tau$ is necessarily non-zero, the equation is not algebraizable using the earlier set of generators~(\ref{jpjzjm}) for which $\tau$ is always equal to $0$. Another importance of non-zero value of $\tau$, which is in terms of quasi-polynomial solutions, is discussed in section~\ref{subsec_heun_polynomials}.
			\item[$\bullet$] \textbf{$\tau$ is a free parameter:}
			This is achieved when $a_6=a_9=0$ in addition to $a_2=0$, i.e., the $O^-$ part is absent. In that case any value of $\tau$ can be used to algebraize the equation.
		\end{itemize}
		
		It may be taken note of, that the parameter $\tau$, associated with the generator $J^-$, is related to the exponent(s) of the local Frobenius series about $z=0$, the actual exponents being given by $2\tau$ (appendix~(\ref{appendix_imp_connections})). There are two of them (Eq.~(\ref{tau_regular})) when $z=0$ is a regular singular point and one of them (Eq.~(\ref{the_thing_dhe})) when $z=0$ is an irregular singular point and $O^-$ is not absent. When $O^-$ is absent, the singularity at $z=0$ is regular, however, both the exponents of this regular singularity, being related to the eigen-value $-a_8$~(appendix~\ref{appendix_imp_connections}), are free parameters.
		
		\textbf{$z=0$ is an ordinary point:}
		
		This corresponds to $a_3 \neq 0$. There is only one Heun class equation (THE) with this property. The $\tau$ value for this equation comes from the analysis of $O^{--}$ part below.
		
		\item The $O^{--}$ part, given by Eq.~(\ref{omm}), is found to be present (section~\ref{sec_eqwise_analysis}) only in the Tri-confluent Heun equation (THE). It is obviously constructed with $J^-J^-$ which, from Eq.~(\ref{newjm}), is found to consist of three kinds of term: $d^2/dz^2, \ (1/z).(d/dz)$ and $1/z^2$. However, $O^{--}$ consists of only the first term. Hence, for the THE, one can take $\tau=0$ to begin with, which yields
		\begin{equation}
		c_{--}=a_3.
		\end{equation}
		No additional algebraizability condition emerges out of this part.
	
		\item The $O^0$ part (Eq.~\ref{o0}) leads to the following three equations:
		\begin{subequations}\label{a1a4a7}
			\begin{align}
			a_1=& c_{+-}, \label{a1} \\
			a_5=& c_0-2c_{+-}\left(\sigma+\tau\right) \label{a4}\\ \mbox{and} \quad
			a_8=& 	c-c_0\left(\sigma+\tau\right)+2c_{+-}\tau\left(1+2\sigma\right), \label{a7}
			\end{align}
		\end{subequations}
		which can be solved for $c_{+-}$, $c_0$ and $c$ in terms of $a_1$, $a_5$ and $a_8$. The parameter $c$, being related to the eigenvalue $-a_8$, will be determined on solving the eigen-value problem represented by Eqs.~(\ref{heun_as_poly}) and (\ref{the_P's}). No additional algebraizability condition arises out of this part.
	\end{itemize}
	
	\subsection{(Quasi-)exact solvability of the Heun class}\label{subsec_heun_polynomials}
	For the purpose of the present work, a differential equation will be called \emph{exactly solvable} when the eigenspectrum of the corresponding differential operator consists of the countably infinite space of quasi-polynomials of the form $z^{\mu}P_N(z)$ (where $\mu \in \mathbb{C}$ and $P_N(z)$ is polynomial of order $N$, $\forall N \in \mathbb{N}$). An equation will be called \emph{quasi-exactly solvable} when only a part of the eigenspectrum consists of quasi-polynomials of the above mentioned form, i.e., the equation admits the above type of solutions for a finite number of $N$. \footnote{It may be noted that Ref.~\cite{turbiner_big} defines (quasi-)exact solvability of a differential equation with respect to the admittance of finite or infinite number of polynomial solutions by the equation. However, an equation admitting solutions of the form $z^{\mu}P_N(z)$ is equivalent to another equation, connected with the previous equation by the F-homotopoic transformation $y(z)=z^{\mu} \xi(z)$, where the equation $\xi(z)$ admits solutions of the form $P_N(z)$. So here the extended definition for (quasi-)exact solvability, involving the quasi-polynomial solutions, is assumed.}
	
	\textbf{Quasi exact solvability:} Once the parameters $\sigma$ and $\tau$ are found in terms of the equation parameters, the quantity $(\sigma-\tau)=j$ (say), referred to hereafter as the \emph{representation parameter}, immediately decides the $su(1,1)$ representation spaces that are available as the solution spaces for the equation in concern (see appendix \ref{appendix_rep_theory}). Since the quasi-polynomial type solutions are of particular interest for the present work, one must look for the conditions that are required for an equation to admit such solutions. For
	\begin{equation}\label{quasi_exact_condn}
	(\sigma-\tau)=j=N/2, \ \mbox{for some} \ N \in \mathbb{N}_0;
	\end{equation}
	the finite $(N+1)$ dimensional representation of $su(1,1)$ is available as solution space (appendix \ref{appendix_rep_theory}), which implies that the corresponding Heun class operator has a finite $(N+1)$ dimensional invariant subspace in the monomial basis. The highest and the lowest states are evidently the states killed by $J^+$ and $J^-$ respectively, which are the monomials $z^{2\sigma}$ and $z^{2\tau}$ respectively. Since the degrees of the $su(1,1)$ generators are $\pm1$ and $0$, hence, the states of the representation are a set of monomials given by
	\begin{equation}\label{monomials}
	z^{2\tau+p} \quad \mbox{where} \quad p \in \{0,1, \dots, 2j\}.
	\end{equation}
	With proper choice(s) of eigenvalue $-a_8$, $(N+1)$ linearly independent linear combinations of the above monomials are obtained as solutions of the concerned equation. The involved process is nothing but that of solving a (N+1) dimensional matrix eigenvalue problem, where the eigenvectors would give the coefficients of the monomials~(\ref{monomials}). These solutions are quasi-polynomial type solutions of the form
	\begin{equation}\label{heun_polynomial}
	y(z)=z^{2\tau}P_N(z)=z^{2\sigma}P_N(1/z)
	\end{equation}
	(where $P_N(z)$ is polynomial in $z$ of order $N=2j$). The corresponding Heun equation is quasi-exactly solvable since the Heun operator admits the Heun-polynomials of the form~(\ref{heun_polynomial}) for a particular $N$. Eq.~(\ref{quasi_exact_condn}) is the \emph{quasi-exact solvability condition}.
	
	\textbf{Exact solvability:} An interesting situation occurs when for a Heun class equation, the $O^+$ part is absent or both of the $O^-$ and the $O^{--}$ parts are absent, i.e., either, $a_0=a_4=a_7=0$ or, $a_2=a_3=a_6=a_9=0$. For example, when $O^+$ is absent, any value of $\sigma$ can be used to cast the concerned equation (as discussed in section~\ref{subsec_algebraizability}) and the value of $\tau$ is determined by the equation parameters as usual. Hence, knowing the value of $\tau$, one can choose countably infinite values of $\sigma$ such that $(\sigma-\tau)=N/2$, $\forall N \in \mathbb{N}_0$. Thus one obtains countably infinite number of quasi-polynomials of the form $z^{2\tau}P_N(z)$ for a specific $\tau \in \mathbb{C}$ and $\forall N \in \mathbb{N}_0$. Similarly, when $O^-$ and $O^{--}$ are absent, $\tau$ becomes free parameter and a countably infinite set of quasi-polynomials of the form $z^{2\sigma}P_N(1/z)$, for a specific $\sigma \in \mathbb{C}$ and $\forall N \in \mathbb{N}_0$, will be obtained. In these cases, the corresponding Heun equation is said to be exactly solvable. Thus the \emph{condition for exact solvability} is given by
	\begin{equation}\label{exact_condn}
	\mbox{either,} \quad a_0=a_4=a_7=0 \implies \mbox{$\sigma$ is free} \quad \mbox{or,} \quad a_2=a_3=a_6=a_9=0  \implies \mbox{$\tau$ is free}.
	\end{equation}
	A more special case of exact solvability occurs when all three of $O^+$, $O^-$ and $O^{--}$ parts are absent in some Heun class equation. Due to the allowed freedom in the choice of both $\sigma$ and $\tau$, the algebra allows an uncountably infinite number of quasi-polynomials of the form $z^{2\tau}P_N(z)$, $\forall \tau \in \mathbb{C}$ and $\forall N \in \mathbb{N}_0$, as solutions. However, since the Heun operator reduces to a diagonal operator in the monomial basis in this case, hence the admitted quasi-polynomial solutions are, in fact, the set of all monomials of the form $\{z^\alpha; \alpha \in \mathbb{C}\}$.
	
	\textbf{Non-zero $\tau$ and additional quasi-polynomials:} For both quasi-exact and exact solvability, the solutions of the form~(\ref{heun_polynomial}) may be associated with zero or non-zero $\tau$ depending on the equation parameters. Now for the generators~(\ref{jpjzjm}), $\tau=0$. The new set of generators, however, besides detecting all the possible cases of $\tau=0$, also opens up the possibilities of finding non-zero $\tau$ value wherever admitted by the equation (appendix~\ref{appendix_imp_connections}). So when $(\sigma-\tau)=N/2 \ (N \in \mathbb{N}_0)$ is associated with some non-zero value of $\tau$, clearly one obtains \emph{additional} quasi-polynomials that can not be directly tracked (see, however, appendix~\ref{subsec_appB_Tur_connection}) using the generators~(\ref{jpjzjm}). The next section focuses on the application of the general algebraic techniques discussed above to the individual equations of Heun class.

	\section{The different equations of the Heun class and quasi-polynomials}\label{sec_eqwise_analysis}
	
	This section deals with the algebraic structure of the individual equations of the Heun class. It is observed that in addition to the polynomials obtained with the earlier set of generators, the new set of generators may lead to additional quasi-polynomial solutions. The algebraizability and (quasi-)exact solvability conditions are provided explicitly for the individual equations.
	
	\subsection{The General Heun Equation (GHE)}\label{subsec_ghe}
	The General Heun equation in its canonical form is given by~\cite{ronveauxbook}
	\begin{equation}\label{ghe}
	\frac{d^2y(z)}{dz^2}+\left(\frac{\gamma}{z}+\frac{\delta}{z-1}+\frac{\varepsilon}{z-a}\right)\frac{dy(z)}{dz}+\frac{\alpha\beta
		z-q}{z(z-1)(z-a)}y(z)=0,
	\end{equation}
	with regular singularities at $z$ = $0, \ 1, \ a(\ne 0,1) \mbox{ and} \ \infty$.
	The exponents at these singularities are $(0, \ 1-\gamma)$, $(0,\ 1-\delta)$, 
	$(0, \ 1-\varepsilon)$ and $(\alpha, \ \beta)$, respectively and $q$ is the eigen-parameter.	Rewriting the above as Eq.~(\ref{heun_as_poly}) one obtains
	\begin{subequations}\label{ghe_a's}
		\begin{align}
		a_0 =& 1 \qquad \qquad \qquad a_4 = 1+\alpha+\beta \qquad \qquad \qquad \qquad
		\qquad a_7 = \alpha\beta \label{ghe_a_op} \\
		a_1 =& -(a+1) \hspace{.85cm} a_5 = -[a\gamma+a\delta-\delta+\alpha+\beta+1]
		\hspace{.93cm} a_8 = -q \label{ghe_a_oz}\\
		a_2 =& a \qquad \qquad \qquad a_6 =  a\gamma \qquad \qquad \qquad \qquad \qquad \qquad \quad a_9=0. \label{ghe_a_om}\\
		a_3=& 0 \label{ghe_a_omm}
		\end{align}
	\end{subequations}
	\begin{description}
		\item[Algebraizability:] From Eqs.~(\ref{ghe_a's}) it is observed that $a_0(=1)$ and $a_2(=a)$ are necessarily non-zero. Hence, the non-algebraizability conditions~(\ref{non_alg_frm_op}) and (\ref{non_alg_frm_om}) are never satisfied, or in other words, GHE is \emph{always algebraizable}. Eq.~(\ref{ghe_a_op}) represents the $O^+$ part. For the GHE, the singularity at $z=\infty$ is regular. Hence, one obtains $\sigma$ from Eq.~(\ref{ghe_sigma}):
		\begin{equation}\label{sigma_ghe}
		\{\sigma_1,\sigma_2\}= \{-\alpha/2, -\beta/2\}.
		\end{equation}
		Eq.~(\ref{ghe_a_om}) represents the $O^-$ part. The GHE has regular singularity at $z=0$. Hence, $\tau$ will be given by Eq.~(\ref{tau_regular}:
		\begin{equation}\label{tau_ghe}
		\{\tau_1,\tau_2\}=\{0, (1-\gamma)/2\}.
		\end{equation}
		\item[Quasi-exact solvability:] Thus four pairs of $\{\sigma,\tau\}$ are obtained, viz., $\{\sigma_1,\tau_1\}$, $\{\sigma_2,\tau_2\}$, $\{\sigma_1,\tau_2\}$ and $\{\sigma_2,\tau_1\}$. If for any one or more of them $(\sigma-\tau)$ is a non-negative half-integer $N/2 \ \mbox{for some} \ N \in \mathbb{N}_0$, the quasi-exact solvability condition~(\ref{quasi_exact_condn}) is satisfied and $(N+1)$ linearly independent (quasi-)polynomial solutions are obtained by solving a $(N+1)$ dimensional matrix eigenvalue problem as discussed in section~\ref{subsec_heun_polynomials}. The subcases are as below:
		\begin{enumerate}
			\item \textbf{Polynomial solutions:} When any of the pairs $\{\sigma_1,\tau_1\}$ and $\{\sigma_2,\tau_1\}$ satisfies condition~(\ref{quasi_exact_condn}), i.e.,
			\begin{equation}
			\alpha=-N \quad \mbox{or} \quad \beta=-N \quad \mbox{for some} \ N \in \mathbb{N}_0,
			\end{equation}
			$(N+1)$ linearly independent polynomial solutions of the form
			\begin{equation}
			\begin{array}{lcl}
			z^{2\tau_1}P_N(z)=P_N(z)=\sum_{i=0}^{N}k_iz^i
			\end{array}
			\end{equation}
			are obtained  for $(N+1)$ values of the eigen-parameter. $-a_8=q$. The coefficients $k_i$ and the eigenvalues are given by the eigenvectors and eigenvalues of the $(N+1)\times(N+1)$ tridiagonal matrix~(\ref{GHE_tridiag_poly}), representing the Heun operator $(\mathcal{H}^{p}_{GHE})$
			in this case, which is discussed in Ref.~\cite{ciftci_hall_saad_dogu} and provided in appendix~C for completeness.
			\item \textbf{Additional quasi-polynomials:} If $\tau_2 \neq 0$ and any of the pairs $\{\sigma_1,\tau_2\}$ and $\{\sigma_2,\tau_2\}$ satisfies condition~(\ref{quasi_exact_condn}), i.e.,
			\begin{equation}\label{GHE_quasi_poly_cond}
			\alpha=(\gamma-1-N) \quad \mbox{or} \quad  \beta=(\gamma-1-N) \quad \mbox{for some} \ N \in \mathbb{N}_0,
			\end{equation}
			$(N+1)$ linearly independent quasi-polynomial solutions of the form
			\begin{equation}
			\begin{array}{lcl}
			z^{2\tau_2}P_N(z)=z^{1-\gamma}P_N(z)=z^{1-\gamma}\sum_{i=0}^{N}k_iz^i
			\end{array}
			\end{equation}
			are obtained for $(N+1)$ values of the eigen-parameter. $-a_8=q$. If the first of the two conditions given by Eq.~(\ref{GHE_quasi_poly_cond}) is satisfied, the coefficients $k_i$ and the eigenvalues are given by the eigenvectors and eigenvalues of the $(N+1)\times(N+1)$ tridiagonal matrix~(\ref{GHE_tridiag_quasi}) representing the Heun operator $(\mathcal{H}^{q}_{GHE})$ in this case. If the second condition given by Eq.~(\ref{GHE_quasi_poly_cond}) is satisfied, the eigenvalue problem to be solved is of a matrix obtained from~(\ref{GHE_tridiag_quasi}) by replacing $\beta$ with $\alpha$. These quasi-polynomial solutions are additional solutions obtained due to the use of generators~(\ref{newgenerators}).
		\end{enumerate}
		\item[Exact solvability:] For GHE one has necessarily non-zero values for $a_0(=1)$ and $a_2(=a)$. Thus the exact-solavbility condition~(\ref{exact_condn}) is never satisfied implying that GHE is \emph{not exactly solvable}.
	\end{description}
	
	\afterpage{
	\begin{landscape}
		\begin{equation}\label{GHE_tridiag_quasi}
		\mathcal{H}^{q}_{GHE}=
		\begin{bmatrix}[1.8]
		\substack{(N-0)(1+0)-(N-0)\gamma\\-\gamma\delta +\beta\gamma+a\gamma\delta-\beta-\delta(a-1)}
		&\substack{2a-a\gamma}&0&0&\dots&0&0&0\\
		
		\substack{N(\gamma-\beta)-N.1}&\substack{(N-1)(1+1)-(N-1)\gamma\\-\gamma\delta +\beta\gamma+a\gamma\delta-2\beta\\-2\delta(a-1)+a\gamma-2a}&\substack{6a-2a\gamma}&0&\dots&0&0&0\\
		
		0&\substack{(N-1)(\gamma-\beta)-(N-1).2}&\substack{(N-2)(1+2)-(N-2)\gamma\\-\gamma\delta +\beta\gamma+a\gamma\delta-3\beta\\-3\delta(a-1)+2a\gamma-6a}&\substack{12a-3a\gamma}&\dots&0&0&0\\
		
		\vdots&\vdots&\vdots&\vdots&\vdots&\vdots&\vdots&\vdots\\
		
		0&0&0&0&\dots&\substack{2(\gamma-\beta)\\-2.(N-1)}&\substack{1.N-\gamma-\gamma\delta +\beta\gamma\\+a\gamma\delta-N\beta-N\delta(a-1)\\+(N-1)a\gamma-N(N-1)a}&\substack{(N+1)(N+2)a\\-(N+1)a\gamma}\\
		
		0&0&0&0&\dots&0&\substack{(\gamma-\beta)-1.N}&\substack{-\gamma\delta+\beta\gamma+a\gamma\delta\\-(N+1)\beta-(N+1)\delta(a-1)\\+Na\gamma-N(N+1)a}
		\end{bmatrix}
		\end{equation}
		
		\vspace{.1cm}
		
		\begin{equation}\label{CHE_tridiag_quasi}
		\mathcal{H}^{q}_{CHE}=
		\begin{bmatrix}[1.8]
		\substack{(1-\gamma)(\delta-\kappa)}
		&\substack{\gamma-2}&0&0&\dots&0&0&0\\
		
		\substack{-N\kappa}&\substack{(2-\gamma)(\delta-\kappa)\\-\gamma+2}&\substack{2\gamma-6}&0&\dots&0&0&0\\
		
		0&\substack{-(N-1)\kappa}&\substack{(3-\gamma)(\delta-\kappa)\\-2\gamma+6}&\substack{3\gamma-12}&\dots&0&0&0\\
		
		\vdots&\vdots&\vdots&\vdots&\vdots&\vdots&\vdots&\vdots\\
		
		0&0&0&0&\dots&\substack{-2\kappa}&\substack{(N-\gamma)(\delta-\kappa)\\-(N-1)\gamma+N(N-1)}&\substack{(N+1)\gamma\\-(N+1)(N+2)}\\
		
		0&0&0&0&\dots&0&\substack{-\kappa}&\substack{(N+1-\gamma)(\delta-\kappa)\\-N\gamma+N(N+1)}
		\end{bmatrix}
		\end{equation}
		
		\vspace{.1cm}
		
		\begin{equation}\label{BHE_tridiag_quasi}
		\mathcal{H}^{q}_{BHE}=
		\begin{bmatrix}[1]
		\substack{\alpha\beta}
		&\substack{1-\alpha}&0&0&\dots&0&0&0\\
		
		\substack{2N}&\substack{\alpha\beta-\beta}&\substack{4-2\alpha}&0&\dots&0&0&0\\
		
		0&\substack{2(N-1)}&\substack{\alpha\beta-2\beta}&\substack{9-3\alpha}&\dots&0&0&0\\
		
		\vdots&\vdots&\vdots&\vdots&\vdots&\vdots&\vdots&\vdots\\
		
		0&0&0&0&\dots&\substack{2.2}&\substack{\alpha\beta-(N-1)\beta}&\substack{(N+1)^2-(N+1)\alpha}\\
		
		0&0&0&0&\dots&0&\substack{2.1}&\substack{\alpha\beta-N\beta}
		\end{bmatrix}
		\end{equation}
	\end{landscape}
}
		
	\subsection{The Confluent Heun Equation (CHE)}\label{subsec_che}
	The Confluent Heun equation is obtained by the coalescence of the regular singularities at $z=a$ (assuming $a>1$, if not the equation should be properly scaled) of the GHE, with the one at infinity. The equation in its canonical form reads~\cite{ronveauxbook}
	\begin{equation}\label{che}
	\dfrac{d^2y(z)}{dz^2} + \left(\kappa+\frac{\gamma}{z}+\frac{\delta}{z-1}\right) \dfrac{dy(z)}{dz}+ \left(\dfrac{\mu}{z}+\dfrac{\nu}{z-1}\right) y(z)=0
	\end{equation}
	with regular singularities at $0$ and $1$ and irregular singularity at infinity. The exponents at the regular singularities are $(0, 1-\gamma)$ and $(0, 1-\delta)$ respectively. Rewriting as Eq.~(\ref{heun_as_poly}) the $a_i$'s are obtained as
	\begin{subequations}\label{che_a's}
		\begin{align}
		a_0 =& 0 \qquad \qquad \qquad a_4 = \kappa \qquad \qquad \qquad \qquad
		\qquad a_7 = \mu+\nu \label{che_a_op} \\
		a_1 =& 1 \hspace{.9cm} \qquad \quad \hspace{.25cm} a_5 = \gamma+\delta-\kappa
		\hspace{1.05cm} \qquad \quad \hspace{.3cm}  a_8 = -\mu \label{che_a_oz}\\
		a_2 =&-1 \qquad \qquad \hspace{.3cm} a_6 =  -\gamma  \qquad \qquad \qquad \qquad \quad \hspace{.1cm} a_9=0. \label{che_a_om} \\
		a_3=& 0 \label{che_a_omm}
		\end{align}
	\end{subequations}
	\begin{description}
		\item[Algebraizability:] Eq.~(\ref{che_a_op}) provides the parameters of $O^+$. Observation shows that the CHE \emph{may be non-algebraizable} for the rare parametric situation characterized by $a_4=\kappa=0$ and $a_7=\mu+\nu \neq 0$, when the non-algebraizability condition~(\ref{non_alg_frm_op}) is fulfilled. For all the other cases when the condition~(\ref{non_alg_frm_op}) is not satisfied, the CHE is algebraizable. Since $z=\infty$ is an irregular singular point $(a_0=0)$ for the CHE, hence two different situations of algebraizability may arise depending on the values of $a_4$ and $a_7$, as discussed in section~\ref{subsec_algebraizability}.
		\begin{itemize}
			\item \textbf{$\sigma$ has a specific value:} For $a_4=\kappa \neq 0$, $\sigma$ is given by (see Eq.~\ref{sigma_cnflnt})
			\begin{equation}\label{sigma_che}
			\sigma=-(\mu+\nu)/2\kappa.
			\end{equation}
			\item \textbf{$\sigma$ is a free parameter:} For $a_4=\kappa=0$ and $a_7=(\mu+\nu)=0$, the original equation does not have the $O^+$ part and thus $\sigma$ becomes a free parameter.
		\end{itemize}
			
		Eq.~(\ref{che_a_om}) gives the parameters of $O^-$. The non-algebraizability condition~(\ref{non_alg_frm_om}) is never fulfilled by CHE, since $a_2(=-1)$ is necessarily non-zero. Hence, no situation of non-algebraizability emerges out of $O^-$. CHE has regular singularity at $z=0$. Hence, $\tau$ will be given by Eq.~(\ref{tau_regular} and the values are
		\begin{equation}\label{tau_che}
		\{\tau_1,\tau_2\}=\{0,(1-\gamma)/2\}.
		\end{equation}
	
		\item[Quasi-exact solvability:] When $\sigma$ is given by Eq.~(\ref{sigma_che}) and $(\sigma-\tau)=j=N/2 \ (\mbox{for some} \ N \in \mathbb{N}_0)$ for any of the $\tau$ value given by Eq.~(\ref{tau_che}), then the condition~(\ref{quasi_exact_condn}) for quasi-exact solvability is satisfied and $(N+1)$ linearly independent (quasi-)polynomials are obtained. The two subcases are discussed below:
		\begin{enumerate}
			\item \textbf{Polynomial solutions:} If the $\{\sigma,\tau_1\}$ pair satisfies the quasi-exact solvability condition, i.e.,
			\begin{eqnarray}
			-(\mu+\nu)/\kappa=N \quad \mbox{for some} \ N \in \mathbb{N}_0,
			\end{eqnarray}
			$(N+1)$ linearly independent polynomial solutions of the form
			\begin{equation}
			\begin{array}{lcl}
			z^{2\tau_1}P_N(z)=P_N(z)=\sum_{i=0}^{N}k_iz^i
			\end{array}
			\end{equation}
			are obtained for $(N+1)$ values of the eigen-parameter. $-a_8=\mu$. The coefficients $k_i$ and the eigenvalues are given by the eigenvectors and eigenvalues of the $(N+1)\times(N+1)$ tridiagonal matrix~(\ref{CHE_tridiag_poly})~\cite{ciftci_hall_saad_dogu}, representing the Heun operator $(\mathcal{H}^{p}_{CHE})$
			in this case.
			\item \textbf{Additional quasi-polynomials:} If $\tau_2 \neq 0$ and the $\{\sigma,\tau_2\}$ pair satisfies the quasi-exact solvability condition, i.e.,
			\begin{eqnarray}
			-(\mu+\nu)/\kappa-(1-\gamma)=N \quad \mbox{for some} \ N \in \mathbb{N}_0,
			\end{eqnarray}
			$(N+1)$ linearly independent quasi-polynomial solutions of the form
			\begin{equation}
			\begin{array}{lcl}
			z^{2\tau_2}P_N(z)=z^{1-\gamma}P_N(z)=z^{1-\gamma}\sum_{i=0}^{N}k_iz^i
			\end{array}
			\end{equation}
			are obtained for $(N+1)$ values of the eigen-parameter. $-a_8=\mu$. The coefficients $k_i$ and the eigenvalues are given by the eigenvectors and eigenvalues of the $(N+1)\times(N+1)$ tridiagonal matrix~(\ref{CHE_tridiag_quasi}) representing the Heun operator $(\mathcal{H}^{q}_{CHE})$ in this case.
		\end{enumerate}
		\item[Exact solvability:] When the equation admits $\sigma$ to be a free parameter, the exact solvability condition~(\ref{exact_condn}) is satisfied. In this case (i.e., with $a_4=\kappa=0$ and $a_7=(\mu+\nu)=0$), the CHE reduces to
		\begin{equation}\label{hypergeometric}
		\dfrac{d^2y(z)}{dz^2} + \left(\frac{\gamma}{z}+\frac{\delta}{z-1}\right) \dfrac{dy(z)}{dz}- \dfrac{\mu}{z(z-1)} y(z)=0.
		\end{equation}
		This is the Hypergeometric equation, which is known to be exactly solvable and here the same fact is established using $su(1,1)$ algebra. The two subcases of exact solvability are:
		\begin{enumerate}
			\item \textbf{Polynomial solutions:} Obtained for the countably infinite set of conveniently chosen $\sigma$, such that the exact solvability condition~(\ref{exact_condn}) is satisfied by the $\{\sigma,\tau_1\}$ pair, i.e.,
			\begin{equation}
			\sigma=N/2 \quad \forall N \in \mathbb{N}_0.
			\end{equation}
			Solutions of the form 
			\begin{equation}
			\begin{array}{lcl}
			z^{2\tau_1}P_N(z)=P_N(z)=\sum_{i=0}^{N}k_iz^i
			\end{array}
			\end{equation}
			are obtained $\forall N \in \mathbb{N}_0$, where the coefficients $k_i$ and the corresponding eigenvalues are given by the eigenvectors and eigenvalues of the matrix obtained from matrix~(\ref{CHE_tridiag_poly}) by putting $\kappa=0$.
			\item \textbf{Additional quasi-polynomials:} Obtained for $\tau_2 \neq 0$ and for the countably infinite set of chosen $\sigma$ such that the $\{\sigma,\tau_2\}$ pair satisfies the exact solvability condition~(\ref{exact_condn}), i.e.,
			\begin{equation}
			\sigma=\{N+(1-\gamma)\}/2 \quad \forall N \in \mathbb{N}_0.
			\end{equation}
			Solutions of the form
			\begin{equation}
			\begin{array}{lcl}
			z^{2\tau_2}P_N(z)=z^{1-\gamma}P_N(z)=z^{1-\gamma}\sum_{i=0}^{N}k_iz^i
			\end{array}
			\end{equation}
			are obtained $\forall N \in \mathbb{N}_0$, where the coefficients $k_i$ and the corresponding eigenvalues are given by the eigenvectors and eigenvalues of the matrix obtained from matrix~(\ref{CHE_tridiag_quasi}) by putting $\kappa=0$.
		\end{enumerate}
	\end{description}
	
	\subsection{The Bi-confluent Heun Equation (BHE)}\label{subsec_bhe}
	The Bi-confluent Heun equation is obtained by the coalescence of the two regular singularities at $z=1$ and $a$ of the GHE with the one at infinity. The canonical form of the equation is~\cite{ronveauxbook}
	\begin{equation}\label{bhe}
	\dfrac{d^2y(z)}{dz^2}+ \left(\dfrac{1+\alpha}{z}-\beta-2z\right)\dfrac{dy(z)}{dz} + \left[\left(\gamma-\alpha-2\right) -\dfrac{1}{2z}\left(\delta+\left(1+\alpha\right)\beta\right)\right]y(z)=0
	\end{equation}
	with regular singularity at $0$ and irregular singularity at $\infty$.
	
	Rewriting the above as Eq.~(\ref{heun_as_poly})
	\begin{subequations}\label{bhe_a's}
		\begin{align}
		a_0 =& 0 \qquad \qquad \qquad a_4 = -2 \qquad \qquad 
		\qquad a_7 = \gamma- \alpha -2 \label{bhe_a_op} \\
		a_1 =& 0 \hspace{.9cm} \qquad \quad \hspace{.25cm} a_5 = -\beta 
		\hspace{1.12cm} \qquad \quad  a_8 = -\dfrac{1}{2}\left(\delta+\left(1+\alpha\right)\beta\right) \label{bhe_a_oz}\\
		a_2 =&1 \qquad \qquad \hspace{.75cm} a_6 =  1+ \alpha \qquad \qquad \quad a_9=0. \label{bhe_a_om}\\
		a_3=&0 \label{bhe_a_omm}
		\end{align}
	\end{subequations}
	\begin{description}
		\item[Algebraizability:] From Eqs.~(\ref{bhe_a's}) it is observed that $a_4(=-2)$ and $a_2(=1)$ are necessarily non-zero. Hence, the non-algebraizability conditions~(\ref{non_alg_frm_op}) and (\ref{non_alg_frm_om}) are never satisfied, or in other words, BHE is \emph{always algebraizable}. The Eq.~(\ref{bhe_a_op}) represents the $O^+$ part. BHE has irregular singularity at $z=\infty$ $(a_0=0)$, however, since $a_4(=-2)$ is strictly non-zero, hence the condition for $\sigma$ being a free parameter is never satisfied (see Eq.~(\ref{exact_condn})), which implies $\sigma$ can be always solved for and is given by (see Eq.~(\ref{sigma_cnflnt}))
		\begin{equation}\label{sigma_bhe}
		\sigma=(\gamma- \alpha -2)/4.
		\end{equation}
		Eq.~(\ref{bhe_a_om}) gives the parameters of $O^-$. The point $z=0$ is a regular singular point for BHE and hence the $\tau$ values will be given by Eq.~(\ref{tau_regular} and the values are
		\begin{equation}\label{tau_bhe}
		\{\tau_1,\tau_2\}=\{0,-\alpha/2\}.
		\end{equation}
		\item[Quasi-exact solvability:] If for any of the two pairs $\{\sigma,\tau_1\}$ and $\{\sigma,\tau_2\}$, the quasi-exact solvability condition~(\ref{quasi_exact_condn}) is satisfied, then $(N+1)$ linearly independent (quasi-)polynomials are obtained. The two subcases are:
		\begin{enumerate}
			\item \textbf{Polynomial solutions:} If the $\{\sigma,\tau_1\}$ pair satisfies the quasi-exact solvability condition, i.e.,
			\begin{equation}
			(\gamma- \alpha -2)/2=N \quad \mbox{for some} \ N \in \mathbb{N}_0,
			\end{equation}
			$(N+1)$ linearly independent polynomial solutions 
			of the form
			\begin{equation}
			\begin{array}{lcl}
			z^{2\tau_1}P_N(z)=P_N(z)=\sum_{i=0}^{N}k_iz^i
			\end{array}
			\end{equation}
			are obtained for $(N+1)$ values of the eigen-parameter. $-a_8= (1/2).\{\delta+(1+\alpha)\beta\}$. The coefficients $k_i$ and the eigenvalues are given by the eigenvectors and eigenvalues of the $(N+1)\times(N+1)$ tridiagonal matrix~(\ref{BHE_tridiag_poly})~\cite{ciftci_hall_saad_dogu}, representing the Heun operator $(\mathcal{H}^{p}_{BHE})$
			in this case.
			\item \textbf{Additional quasi-polynomials:} If $\tau_2 \neq 0$ and the $\{\sigma,\tau_2\}$ pair satisfies the quasi-exact solvability condition, i.e.,
			\begin{equation}
			(\gamma- \alpha -2)/2+\alpha=(\gamma+ \alpha -2)/2=N \quad \mbox{for some} \ N \in \mathbb{N}_0,
			\end{equation}
			$(N+1)$ linearly independent quasi-polynomial solutions of the form
			\begin{equation}
			\begin{array}{lcl}
			z^{2\tau_2}P_N(z)=z^{-\alpha}P_N(z)=z^{-\alpha}\sum_{i=0}^{N}k_iz^i
			\end{array}
			\end{equation}
			are obtained for $(N+1)$ values of the eigen-parameter. $-a_8=(1/2).\{\delta+(1+\alpha)\beta\}$. The coefficients $k_i$ and the eigenvalues are given by the eigenvectors and eigenvalues of the $(N+1)\times(N+1)$ tridiagonal matrix~(\ref{BHE_tridiag_quasi}), representing the Heun operator $(\mathcal{H}^{q}_{BHE})$
			in this case.
		\end{enumerate}
		\item[Exact solvability] For BHE one has necessarily non-zero values for $a_4(=-2)$ and $a_2(=1)$. Thus the exact-solavbility condition~(\ref{exact_condn}) is never satisfied implying that BHE is \emph{not exactly solvable}.
	\end{description}
	
	\subsection{The Doubly-confluent Heun Equation (DHE)}\label{subsec_dhe}
	
	The Doubly-confluent Heun equation is obtained from the GHE by coalescing one of its two regular singularities at $z=1$ and $a$ with the regular singularity at infinity and the other with the regular singularity at $0$. The canonical form of the equation is given by~\cite{ronveauxbook}
	\begin{eqnarray}\label{dhe}
	\nonumber z^2\dfrac{d^2y(z)}{dz^2}+ \left(\alpha_1z^2+z+\alpha_{-1}\right)\dfrac{dy(z)}{dz} \hspace{4.75cm}\\
	+\left(\left(B_1+\dfrac{\alpha_1}{2}\right)z+\left(B_0+\dfrac{\alpha_1\alpha_{-1}}{2}\right)+\left(B_{-1}-\dfrac{\alpha_{-1}}{2}\right)\dfrac{1}{z}\right)y(z)=0
	\end{eqnarray}
	with irregular singularities at both $0$ and $\infty$. The parameters are given by	
	\begin{subequations}\label{dhe_a's}
		\begin{align}
		a_0 =& 0 \qquad \qquad \qquad a_4 = \alpha_1 \qquad \qquad 
		\qquad a_7 = B_1+ \alpha_1/2 \label{dhe_a_op} \\
		a_1 =& 1 \hspace{.9cm} \qquad \quad \hspace{.25cm} a_5 = 1 
		\hspace{1.38cm} \qquad \quad  a_8 = B_0+ \alpha_1\alpha_{-1}/2 \label{dhe_a_oz}\\
		a_2 =&0 \qquad \qquad \hspace{.75cm} a_6 =\alpha_{-1} \qquad \qquad \quad \hspace{.16cm} a_9=B_{-1}- \alpha_{-1}/2. \label{dhe_a_om}\\
		a_3=&0 \label{dhe_a_omm}
		\end{align}
	\end{subequations}

	\afterpage{
		\begin{landscape}
			\begin{equation}\label{DHE_tridiag_quasi}
			\mathcal{H}^{q(i)}_{DHE}=
			\begin{bmatrix}[2]
			\substack{1/4+B^2_{-1}/\alpha^2_{-1}\\-B_{-1}/\alpha_{-1}}
			&\substack{\alpha_{-1}}&0&0&\dots&0&0&0\\
			
			\substack{-N\alpha_1}&\substack{9/4+B^2_{-1}/\alpha^2_{-1}\\-3B_{-1}/\alpha_{-1}}&\substack{2\alpha_{-1}}&0&\dots&0&0&0\\
			
			0&\substack{-(N-1)\alpha_1}&\substack{25/4+B^2_{-1}/\alpha^2_{-1}\\-5B_{-1}/\alpha_{-1}}&\substack{3\alpha_{-1}}&\dots&0&0&0\\
			
			\vdots&\vdots&\vdots&\vdots&\vdots&\vdots&\vdots&\vdots\\
			
			0&0&0&0&\dots&\substack{-2\alpha_1}&\substack{(2N-1)^2/4+B^2_{-1}/\alpha^2_{-1}\\-(2N-1)B_{-1}/\alpha_{-1}}&\substack{(N+1)\alpha_{-1}}\\
			
			0&0&0&0&\dots&0&\substack{-\alpha_1}&\substack{(2N+1)^2/4+B^2_{-1}/\alpha^2_{-1}\\-(2N+1)B_{-1}/\alpha_{-1}}
			\end{bmatrix}
			\end{equation}
			
			\vspace{3.5cm}
			
			\begin{equation}\label{DHE_tridiag_quasi_2}
			\mathcal{H}^{q(ii)}_{DHE}=
			\begin{bmatrix}[2]
			\substack{1/4+B^2_{1}/\alpha^2_{1}\\+B_{1}/\alpha_{1}}
			&\substack{-\alpha_{1}}&0&0&\dots&0&0&0\\
			
			\substack{N\alpha_{-1}}&\substack{9/4+B^2_{1}/\alpha^2_{1}\\+3B_{1}/\alpha_{1}}&\substack{-2\alpha_{1}}&0&\dots&0&0&0\\
			
			0&\substack{(N-1)\alpha_{-1}}&\substack{25/4+B^2_{1}/\alpha^2_{1}\\+5B_{1}/\alpha_{1}}&\substack{-3\alpha_{1}}&\dots&0&0&0\\
			
			\vdots&\vdots&\vdots&\vdots&\vdots&\vdots&\vdots&\vdots\\
			
			0&0&0&0&\dots&\substack{2\alpha_{-1}}&\substack{B^2_{1}/\alpha^2_{1}+(2N-1)^2/4\\+(2N-1)B_{1}/\alpha_{1}}&\substack{-(N+1)\alpha_{1}}\\
			
			0&0&0&0&\dots&0&\substack{\alpha_{-1}}&\substack{(2N+1)^2/4+B^2_{1}/\alpha^2_{1}\\+(2N+1)B_{1}/\alpha_{1}}
			\end{bmatrix}
			\end{equation}
		\end{landscape}
	}
	
	\begin{description}
		\item[Algebraizability:] Eq.~(\ref{dhe_a_op}) provides the parameters of $O^+$. Observation shows that \emph{non-algebraizability may arise} for the DHE for the rare situation characterized by $a_4=\alpha_1=0$ but $B_1 \neq 0$ (implying $a_7=B_1+\alpha_1/2 \neq 0$), when the non-algebraizability condition~(\ref{non_alg_frm_op}) is fulfilled. For all the other cases when the condition~(\ref{non_alg_frm_op}) is not satisfied, the $O^+$ part is algebraizable. Since $z=\infty$ is an irregular singular point $(a_0=0)$ for the DHE, hence two different situations of algebraizability may arise depending on the values of $a_4$ and $a_7$, as discussed in section~\ref{subsec_algebraizability}.
		\begin{itemize}
			\item \textbf{$\sigma$ has a specific value:} For $a_4=\alpha_1 \neq 0$, $\sigma$ is given by (see Eq.~\ref{sigma_cnflnt})
			\begin{equation}\label{sigma_dhe}
			\sigma=-(B_1/2\alpha_1+1/4)
			\end{equation}
			\item \textbf{$\sigma$ is a free parameter:} For $a_4=\alpha_1=0$ and $a_7=B_1+\alpha_1/2=0$, the original equation does not have the $O^+$ part and thus $\sigma$ becomes a free parameter.
		\end{itemize}
		Eq.~(\ref{dhe_a_om}) gives the parameters of $O^-$. Observation shows that DHE \emph{may be non-algebraizable} for the rare situation characterized by $a_6=\alpha_{-1}=0$ but $B_{-1} \neq 0$ (such that $a_7=B_{-1}+\alpha_{-1}/2 \neq 0$), when the non-algebraizability condition~(\ref{non_alg_frm_om}) is fulfilled. For all the other cases when the condition~(\ref{non_alg_frm_om}) is not satisfied, the $O^-$ part is algebraizable. Since $z=0$ is an irregular singular point $(a_2=a_3=0)$ for the DHE, hence two different situations of algebraizability may arise depending on the values of $a_6$ and $a_9$, as discussed in section~\ref{subsec_algebraizability}.
		\begin{itemize}
			\item \textbf{$\tau$ has a specific value:} For $a_6=\alpha_{-1} \neq 0$, $\tau$ is given by (see Eq.~\ref{the_thing_dhe})
			\begin{equation}\label{tau_dhe}
			\tau=-(B_{-1}/2\alpha_{-1}-1/4),
			\end{equation}
			\item \textbf{$\tau$ is a free parameter:} For $a_6=\alpha_{-1}=0$ and $a_7=B_{-1}+\alpha_{-1}/2=0$, the original equation does not have the $O^-$ and $O^{--}$ parts and thus $\tau$ becomes a free parameter.
		\end{itemize}
		It is important to note that unlike all other equations of the Heun class, the DHE does not necessarily have at least one $\tau$ value equal to $0$. Hence, for this equation polynomial solution(s) may only be obtained if either the $\tau$ value given by Eq.~(\ref{tau_dhe}) is $0$ or $\tau$ is a free parameter which can be chosen to be $0$. Evidently, for both of these cases $\sigma$ must either be free or the $\sigma$ value given by Eq.~(\ref{sigma_dhe}) must be a non-negative half-integer. Hence, \emph{for a DHE with more general parameter values, which may not admit polynomial solutions, the quasi-polynomial solutions may prove to be of much importance}. The conditions for obtaining them are listed below for the quasi-exactly and exactly solvable cases. It may also be noted that when the $\tau$ value is given by Eq.~(\ref{tau_dhe}) and it is non-zero, then the equation \emph{can not be algebraized directly} with generators~(\ref{jpjzjm}), for which $\tau$ is necessarily zero. However, with an initial transformation of the dependent variable, algebraization in terms of generators~(\ref{jpjzjm}) is possible (appendix~\ref{subsec_appB_Tur_connection}).
		\item[Quasi-exact solvability:] When $\sigma$ is given by Eq.~(\ref{sigma_dhe}), $\tau$ is given by Eq.~(\ref{tau_dhe}) and the $\{\sigma,\tau\}$ pair satisfies the quasi-exact solvability condition~(\ref{quasi_exact_condn}), i.e.,
		\begin{equation}
		-(B_{1}/\alpha_{1}+1/2)+(B_{-1}/\alpha_{-1}-1/2)=N \quad \mbox{for some} \ N \in \mathbb{N}_0,
		\end{equation}
		$(N+1)$ linearly independent (quasi-)polynomials
		are obtained. These are of the form 
		\begin{equation}
		\begin{array}{lcl}
		z^{2\tau}P_N(z)=P_N(z)=\sum_{i=0}^{N}k_iz^i
		\end{array}
		\end{equation}
		if $\tau=-(B_{-1}/2\alpha_{-1}-1/4)=0$ and are \emph{additional quasi-polynomials} of the form
		\begin{equation}
		\begin{array}{lcl}
		z^{2\tau}P_N(z)=z^{(1/2-B_{-1}/\alpha_{-1})}P_N(z)=z^{(1/2-B_{-1}/\alpha_{-1})}\sum_{i=0}^{N}k_iz^i
		\end{array}
		\end{equation}
		if $\tau \neq 0$. The coefficients $k_i$ and the eigenvalues $-a_8=-B_0-\alpha_1\alpha_{-1}/2$ are given by the eigenvectors and eigenvalues of any of the two equivalent $(N+1)\times(N+1)$ tridiagonal matrix $(\mathcal{H}^{q(i)}_{DHE})$~(\ref{DHE_tridiag_quasi}) or $(\mathcal{H}^{q(ii)}_{DHE})$~(\ref{DHE_tridiag_quasi_2}) representing the Heun operator in this case.
		\item[Exact solvability:] When the equation allows any one of $\sigma$ and $\tau$ to be a free parameter, then the exact solvability condition~(\ref{exact_condn}) is satisfied. The subcases are discussed below:
		\begin{enumerate}
			\item \textbf{$\sigma$ is free, $\tau$ is fixed:} This mean $\alpha_1=B_1=0$, $\tau$ is given by Eq.~(\ref{tau_dhe}) and one is free to choose $\sigma$ from a countably infinite set such that the $\{\sigma,\tau\}$ pair satisfies the exact solvability condition~(\ref{exact_condn}), i.e., 
			\begin{equation}
			\sigma=N/2-(B_{-1}/2\alpha_{-1}-1/4) \quad \forall N \in \mathbb{N}_0.
			\end{equation}
			Solutions of the form
			\begin{equation}
			\begin{array}{lcl}
			z^{2\tau}P_N(z)=P_N(z)=\sum_{i=0}^{N}k_iz^i, \quad \forall N \in \mathbb{N}_0
			\end{array}
			\end{equation}
			are obtained if $\tau=1/4-B_{-1}/\alpha_{-1}=0$ and of the form 
			\begin{equation}
			\begin{array}{lcl}
			z^{2\tau}P_N(z)=z^{(1/2-B_{-1}/\alpha_{-1})}P_N(z)=z^{(1/2-B_{-1}/\alpha_{-1})}\sum_{i=0}^{N}k_iz^i, \quad \forall N \in \mathbb{N}_0
			\end{array}
			\end{equation}
			are obtained if $\tau \neq 0$. The coefficients $k_i$ and the corresponding eigenvalues are given by the eigenvectors and eigenvalues of the matrix obtained from matrix~(\ref{DHE_tridiag_quasi}) by putting $\alpha_1=0$. In this case (i.e., when $a_4=\alpha_1=0$ and also $B_1=0$ implying $a_7=0$) the DHE takes the form
			\begin{equation}\label{dhe_spcl_1}
			z^2\dfrac{d^2y(z)}{dz^2}+ \left(z+\alpha_{-1}\right)\dfrac{dy(z)}{dz}
			+\left(B_0+\left(B_{-1}-\dfrac{\alpha_{-1}}{2}\right)\dfrac{1}{z}\right)y(z)=0,
			\end{equation}
			which is a confluent hypergeometric equation upto the independent variable transformation $t=1/z$. With an additional F-homotopic transformation $t^{\iota B_0}$, the confluent hypergeometric equation is obtained in its canonical form. Thus an equation of the form~(\ref{dhe_spcl_1}) is found to be exactly solvable.
			\item \textbf{$\tau$ is free, $\sigma$ is fixed:} $\sigma$ is given by Eq.~(\ref{sigma_dhe}) and one is free to choose $\tau$ from a countably infinite set such that the $\{\sigma,\tau\}$ pair satisfies the exact solvability condition~(\ref{exact_condn}), i.e., 
			\begin{equation}
			\tau=-N/2-(B_{1}/2\alpha_{1}+1/4) \quad \forall \ N \in \mathbb{N}_0.
			\end{equation}
			Solutions of the form 
			\begin{equation}
			\begin{array}{lcl}
			z^{2\sigma}P_N(1/z)=z^{-(B_1/\alpha_1+1/2)}P_N(1/z)=z^{-(B_1/\alpha_1+1/2)}\sum_{i=0}^{N}k_i({1 \over z})^i, \quad \forall N \in \mathbb{N}_0
			\end{array}
			\end{equation}
			are obtained, where the coefficients $k_i$ and the corresponding eigenvalues are given by the eigenvectors and eigenvalues of the matrix obtained from matrix~(\ref{DHE_tridiag_quasi_2}) by putting $\alpha_{-1}=0$. All these are in general \emph{additional quasi-polynomials} of the form $z^{2\tau}P_N(z)$ (since, $\tau \neq 0$ in general) except when $\sigma \in \mathbb{N}_0$, in which case there will be a single polynomial corresponding to $\tau=0$. In this case (i.e., when $a_6=\alpha_{-1}=0$ and also $B_{-1}=0$ implying $a_9=0$) the DHE takes the form
			\begin{equation}\label{dhe_spcl_2}
			z^2\dfrac{d^2y(z)}{dz^2}+ \left(\alpha_1z^2+z\right)\dfrac{dy(z)}{dz}
			+\left(\left(B_1+\dfrac{\alpha_1}{2}\right)z+B_0\right)y(z)
			=0,
			\end{equation}
			which is a confluent hypergeometric equation, the canonical form of which is obtained with the F-homotopic transformation $z^{\iota B_0}$. Thus the confluent hypergeometric equation is found to be exactly solvable.
			\item \textbf{Both $\sigma$ and $\tau$ are free:} This is a very special case of exact solvability when one has $\alpha_1=B_1=\alpha{-1}=B_{-1}=0$. The DHE reduces to
			\begin{equation}\label{dhe_spcl_3}
			z^2\dfrac{d^2y(z)}{dz^2}+ z\dfrac{dy(z)}{dz}+B_0y(z)=0,
			\end{equation}
			an equation with two singularities at $z=0$ and $z=\infty$, both of which are regular. The differential operator represented by Eq.~(\ref{dhe_spcl_3}) is a diagonal operator and admits all monomial solutions of the form $\{z^c,\ c \in \mathbb{C}\}$, as discussed in section~\ref{subsec_heun_polynomials}. The eigenvalue corresponding to $z^c$ is $-B_0=c^2$, as can be easily seen from the equation.
		\end{enumerate}
	\end{description}
		
	\subsection{The Tri-confluent Heun Equation (THE)}
	The Tri-confluent Heun equation is obtained from the GHE by coalescing all its finite regular singularities with the one at $z=\infty$. The THE in its canonical form is given by~\cite{ronveauxbook}
	\begin{equation}\label{the}
	\dfrac{d^2y(z)}{dz^2}-\left(3z^2+\gamma\right)\dfrac{dy(z)}{dz}+\left[\alpha+(\beta-3)\right]y(z)=0,
	\end{equation}
	with irregular singularity at $z=\infty$. The parameters read
	\begin{subequations}\label{the_a's}
		\begin{align}
		a_0 =& 0 \qquad \qquad \qquad a_4 = -3 \qquad \qquad 
		\quad \hspace{.35cm} a_7 = 0 \label{the_a_op} \\
		a_1 =& 0 \hspace{.9cm} \qquad \quad \hspace{.25cm} a_5 = 0 
		\hspace{1.4cm} \qquad \quad  a_8 = \alpha+(\beta-3) \label{the_a_oz}\\
		a_2 =&0 \qquad \qquad \hspace{.75cm} a_6 =  -\gamma \qquad \qquad \quad \hspace{.3cm} a_9=0. \label{the_a_om}\\
		a_3=&1 \label{the_a_omm}
		\end{align}
	\end{subequations}
	\begin{description}
		\item[Algebraizability:] From Eqs.~(\ref{the_a's}) it is observed that $a_4(=-3)$ is necessarily non-zero and $a_9=0$. Hence, the non-algebraizability conditions~(\ref{non_alg_frm_op}) and (\ref{non_alg_frm_om}) are never satisfied, or in other words, THE is \emph{always algebraizable}.
		
		Here one has $O^{--}$, the degree $-2$ part, which contains a single term $d^2y/dz^2$. This is constructed using $J^-J^-$ with $\tau=0$, which makes the new set of generators identical to the earlier one. Eq.~(\ref{the_a_op}) gives the parameters of $O^+$. THE has irregular singularity at $z=\infty$ $(a_0=0)$, however, since $a_4(=-3)$ is strictly non-zero, hence the condition for $\sigma$ being a free parameter is never satisfied (see Eq.~(\ref{exact_condn})), which implies $\sigma$ can be always solved for and is given by (see Eq.~(\ref{sigma_cnflnt}))
		\begin{equation}
		\sigma=0.
		\end{equation}
		\item[Quasi-exact solvability:] Thus one has $\sigma=\tau=0$ implying $\sigma-\tau=0$ leading to a singlet solution which is $z^{2\tau}P_0(z)=$constant.
		\item[Exact solvability:] Eq.~(\ref{the_a_op}) shows that the exact solvability condition $a_0=a_4=a_7=0$ allowing $\sigma$ to  be a free parameter can never be satisfied, since $a_4(=-3)$ is necessarily non-zero. Eq.~(\ref{the_a_omm}) tells that $a_3 \neq 0$, hence the other condition for exact solvability $a_2=a_3=a_6=a_9=0$ is also not possible for THE. Thus the THE is \emph{not exactly solvable}.
	\end{description}
	
	\section{Conclusion}\label{sec_conclusion}
	
	To summarize, the Heun class of equations has been algebraized using a bi-parametric set of $su(1,1)$ generators of degrees $\pm 1$ and $0$. Algebraization using these new generators incorporates the algebraization due to an earlier set of generators as a subset both in terms of number of algebraizable equations and number of available quasi-polynomials--- a type of global solutions which are of utter importance both mathematically and physically. Conditions for algebraizability, quasi-exact solvability and exact solvability have been listed for generic and equationwise cases. Two conditions for non-algebraizability (which may occur for the confluent and the doubly confluent Heun equations) have been identified which speaks in contrary to the common belief that Heun class of equations must have $sl(2,R)$ structure. For the quasi-exactly solvable and exactly solvable cases, explicit conditions leading to additional quasi-polynomials have been provided for the convenience of direct use by the reader.
	
	The method presented in this paper can be applied to Heun class of equations appearing in innumerable physics problems of gravitational, cosmological, quantum mechanical or material science origin. One such physics problem is that of the massless Dirac particle in the $C$-metric, for which both the radial and polar parts of the Dirac equation takes the form of the general Heun equation~\cite{bini_bit_ger}. Classes of quasi-polynomial solutions to the radial and the polar parts of the Dirac equation have been obtained using the methods presented in this paper and will be presented elsewhere~\cite{third_paper}.
	
	\section*{Acknowledgments:} The author conveys heartfelt thanks to Dr. Ritesh K. Singh and to Dr. Ananda Dasgupta for useful discussions.
	
	\newpage
	
	\appendix
	\begin{center}
		\Huge{Appendices}
	\end{center}
	\section{The $su(1,1)$ representation theory}\label{appendix_rep_theory}

	\renewcommand{\thetable}{\thesection.\arabic{table}}
	\renewcommand{\thefigure}{\thesection.\arabic{figure}}
	
	The $su(1,1)$ algebra may be regarded as a kind of deformation of the $su(2)$ algebra. The representations of a generic deformed $su(2)$ have been extensively studied in ref.~\cite{rocek}. The $su(1,1)$ representation theory is discussed below as a special case of the general deformed $su(2)$ representation theory.
	
	\subsection{Representations of deformed $su(2)$ and $su(1,1)$}\label{subsec_appA_rep_theory_general}
	
\begin{figure}[!ht]
		\centerline{\includegraphics[width=15.302cm]{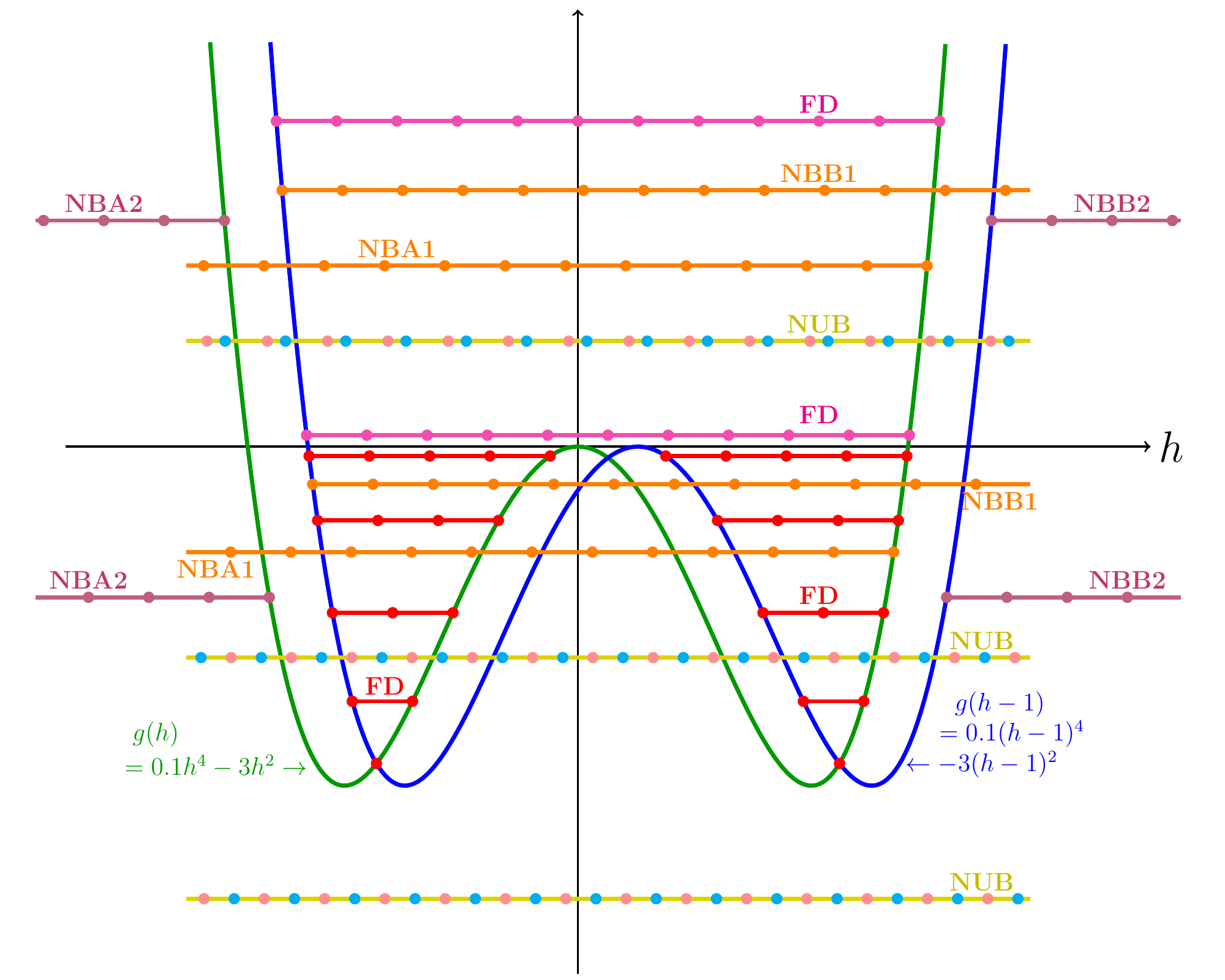}}
		\caption{\label{fig_dummy_deformed_su2}{\emph{Different representations of a general deformed $su(2)$.} NUB: \emph{Non-unitary UnBounded,} NBB1 \emph{and} NBB2: \emph{Non-unitary Bounded Below (types 1 and 2)}, NBA1 \emph{and} NBA2: \emph{Non-unitary Bounded Above (types 1 and 2),} FD: \emph{Finite Dimensional, which are unitary in this particular example.}}}
	\end{figure}

	The general deformed $su(2)$ algebra in the Cartan-Weyl basis is given by\cite{rocek}
	\begin{equation}\label{deformed_su(2)}
	\left[H,E_{\pm}\right]=\pm E_{\pm}, \qquad \left[E_+,E_-\right]=f(H)
	\end{equation}
	with Casimir
	\begin{equation}\label{general_casimir}
	C=E_-E_+ + g(H),
	\end{equation}
	where, $g(H)$ is given by
	\begin{equation}\label{fh_gh}
	f(H)=g(H)-g(H-1).
	\end{equation}
	Here $H$ is the only element of the Cartan subalgebra. For $su(2)$ one has $f(H)=2H$ (and $g(H)=H(H+1)$), whereas, other functional forms of $f(H)$ lead to deformed $su(2)$'s. Plotting the functions $g(h)$ and $g(h-1)$ versus $h$ (where, $h$ is eigenvalue of $H$) makes it convenient to discuss the representation theory\cite{rocek}. The Casimir operator is an invariant of the algebra, i.e., for a specific representation the Casimir is a constant represented by a horizontal line in the $h-g(h)$ plot. The states $|c,h\rangle$ (labeled by the Casimir value $c$ and the $H$ eigenvalue $h$) are represented by points at unit distances along the Casimir line. The operators $E_+$ and $E_-$ act as raising and lowering operators respectively and from the basic properties of raising and lowering coefficients one immediately observes that
	\begin{enumerate}[label=(\roman*)]
		\item  for unitary representations all the states must lie above the $g(h)$ curve and
		\item the highest weight of a bounded above representation must lie on the $g(h)$ curve, whereas the lowest weight of a bounded below representation must lie on the $g(h-1)$ curve. 
	\end{enumerate}
	The finite dimensional representations are special cases of the bounded above and below representations for which the highest and the lowest states must lie on the $g(h)$ and $g(h-1)$ curves respectively. From the construct of the raising-lowering operators, the weights of any representation must differ from one another by integers. For any bounded representation, there is only a single way to select a set of states with the above property (since it has either the highest or the lowest weight or both, fixed)--- they are known as the discrete representations. For any unbounded representation, the valid set of states can be chosen in innumerably infinite ways (since no weight for them is fixed)--- they are known as the continuous representations. Based on these basic rules, the classes of representations of any deformed $su(2)$ algebra can be figured out. The representations of an arbitrary deformation of $su(2)$ are shown in Fig.~\ref{fig_dummy_deformed_su2}.
	
\begin{figure}[!ht]
			\centerline{\includegraphics[width=15.8cm]{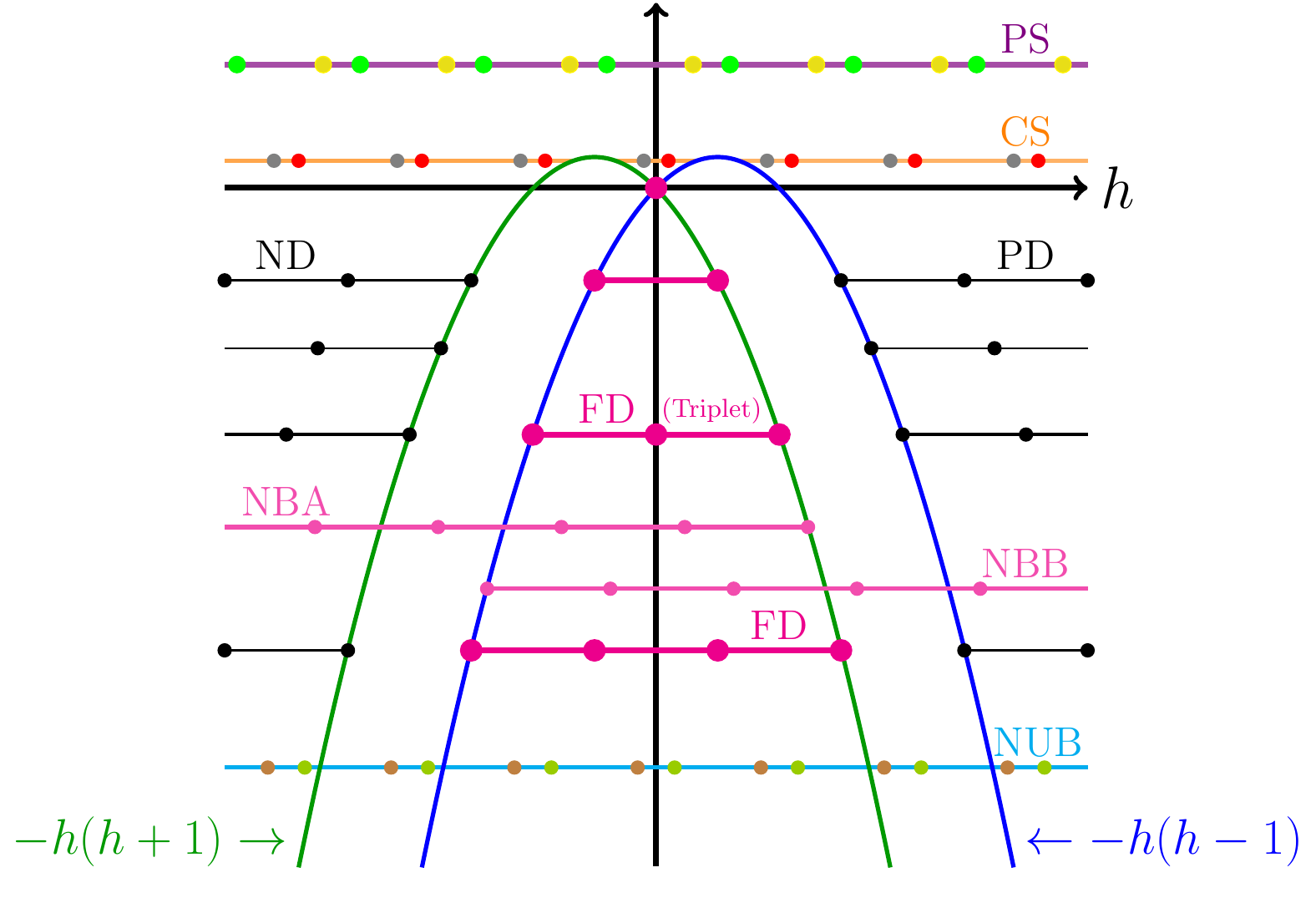}}
			\caption{\label{fig_SU11}{\emph{Different representations of $su(1,1)$. Four finite dimensional representations (singlet to quadruplet) have been shown which are special cases of} NBA \emph{and} NBB \emph{type representations, available for specific Casimir values.} NBA \emph{and} NBB \emph{for general Casimir values are also shown and so are the} PD \emph{and} ND \emph{type representations. Two sets of different colored states for each of the representations} PS, CS \emph{and} NUB \emph{represent two different representations with the same Casimir value, among the uncountably infinite possibilities for a specific Casimir value that are available for these three types of continuous representations.}}}
		\end{figure}

	Now for ${su}(1,1)$,
		\begin{equation}\label{fh}
		f(H)=-2H
		\end{equation}
		and Eq.~(\ref{fh_gh}) yields
		\begin{equation}\label{gh}
		g(H)=-H(H+1).
		\end{equation}
		The Casimir is given by
		\begin{equation}\label{general_casimir0}
		C=E_-E_+ + g(H)=E_-E_+ - H(H+1),
		\end{equation}
		or alternatively (using Eqs.~(\ref{deformed_su(2)}) and (\ref{fh_gh})),
		\begin{equation}
		C=E_+E_- + g(H-1)=E_+E_- - H(H-1).
		\end{equation}
		The plot of $g(h)$ and $g(h-1)$  is shown in Fig.~\ref{fig_SU11}. These are parabolas with their vertices at the points $(-1/2,1/4)$ and $(1/2,1/4)$, respectively. For any Casimir value $c<1/4$, the constant $c$ line intersects the $g(h)$ curve at the points $h=(-1 \pm \sqrt{1-4c})/2$. Based on the general rules for the representations of deformed $su(2)$, the representations of $su(1,1)$ are listed below:
		\begin{description}
			\item[Unitary:] Unitary representations\footnote{See references~\cite{su11_adams,su11_barut,su11_book} for details of unitary representations of $su(1,1)$} of $su(1,1)$ can be divided into two subclasses:
			\begin{enumerate}
				\item \textbf{Unbounded:} There can be two types of unitary unbounded representations:
				\begin{enumerate}
					\item  \textbf{The Principal Series (PS)}:
					
					$c\in [1/4,\infty)$, $h\in(-\infty,\infty)$, $\{c,h\}\neq\{1/4,1/2\}$.
					\item \textbf{The Complementary Series (CS)}:
					
					$c\in
					(0,1/4)$,  $h\in(-\infty,\infty)$, $h\notin \left[\dfrac{-1- \sqrt{1-4c}}{2}, \dfrac{-1+ \sqrt{1-4c}}{2}\right]$.
				\end{enumerate}
				\item  \textbf{Bounded:} Once again two different types of representation spaces are available:
				\begin{enumerate}
					\item \textbf{The Positive Discrete (PD) Series} (Bounded below):
					
					$c\le 1/4$. For $1/4 \ge c >0$,
					$h_{min} = \dfrac{1 \pm \sqrt{1-4c}}{2}$. For $c \le 0$, $h_{min} = \dfrac{1 + \sqrt{1-4c}}{2}$.
					\item \textbf{The Negative Discrete (ND) Series} (Bounded above):
					
					$c\le 1/4$. For $1/4 \ge c >0$, $h_{max} = \dfrac{-1 \pm \sqrt{1-4c}}{2}$. For $c \le 0$, $h_{max} = \dfrac{-1 - \sqrt{1-4c}}{2}$.
				\end{enumerate}
			\end{enumerate}
			\item[Non-Unitary:] Non-unitary representations of $su(1,1)$ can be subdivided into two classes:
			\begin{enumerate}
				\item \textbf{Non-unitary Unbounded (NUB):}
				
				$c\in (-\infty,1/4)$, $h\in(-\infty,\infty)$, $h\neq \dfrac{-1\pm \sqrt{1-4c}}{2}$. $h \in \left(\dfrac{-1- \sqrt{1-4c}}{2}, \dfrac{-1+ \sqrt{1-4c}}{2}\right)$ must hold for at least one $h$.
				\item \textbf{Non-unitary Bounded:} This can be subdivided into three classes:
				\begin{enumerate}
					\item \textbf{Non-unitary Bounded Below (NBB):} $c\le 0$,
					$h_{min} = \dfrac{1 - \sqrt{1-4c}}{2}$.
					\item \textbf{Non-unitary Bounded Above (NBA)}:
					
					$c\le 0$,
					$h_{max} = \dfrac{-1 + \sqrt{1-4c}}{2}$.
					\item \textbf{Bounded both sides or Finite Dimensional (FD)}:
					
					$N \ (\in \mathbb{N})$ dimensional space for $c=(1-N^2)/4$, $h_{min} = \dfrac{1 - \sqrt{1-4c}}{2}=\dfrac{1-N}{2}$, $h_{max} = \dfrac{-1 + \sqrt{1-4c}}{2}=\dfrac{N-1}{2}$.
				\end{enumerate}
			\end{enumerate}
		\end{description}
		
		\begin{table}[!ht]
			\caption{\label{Table_SU11_reps}{Representations of $su(1,1)$ at a glance}}
		\begin{center}
			\begin{tabular} {|p{1.2cm}|p{7cm}|p{7cm}|}
				\hline
				 & Unbounded & Bounded \\ 
				\hline 
				Unitary 
				& \textbf{(a) The Principal Series (PS)}
				
				\textbf{(b) The Complementary Series (CS)}
				
				& \textbf{(a) The Positive Discrete (PD) Series} (Bounded below)
				
				\textbf{(b) The Negative Discrete (ND) Series} (Bounded above)
				
				 \\ 
				\hline 
				Non-unitary 
				& \textbf{Non-unitary Unbounded} (\textbf{NUB}):
				
				& \textbf{(a) Non-unitary Bounded Below (NBB)}
				
				\textbf{(b) Non-unitary Bounded Above (NBA)}
				
				\textbf{(c) Bounded both sides or Finite Dimensional (FD)} \\ 
				\hline
			\end{tabular}
		\end{center}
		\end{table}

	\subsection[Solutions of Heun class using $su(1,1)$ representation theory]{Solutions of Heun class of equations using $su(1,1)$ representation theory}\label{subsec_appA_rep_theory_for_heun}
	
	This section shows how the general $su(1,1)$ representation spaces discussed above serve as the solution spaces for the Heun class of equations. Once the parameters $\sigma$ and $\tau$ are determined by the equation in concern (as discussed in the  section~\ref{sec_our_gens_&_heun_class} for Heun class in general and in the section~\ref{sec_eqwise_analysis} for individual equations), one immediately finds  (i) the Casimir value (given by Eq.~(\ref{construct_casimir})) associated with the representations available as the Heun solutions spaces and (ii) two monomials $z^{2\sigma}$ and $z^{2\tau}$, which are killed by the raising operator and the lowering operator respectively. Thus a bounded above and a bounded below representation of $su(1,1)$ are associated with each equation where the highest and the lowest weight states are proportional to the above two monomials respectively. Hence, the bounded below representation hosts solution of the form
	\begin{equation}\label{heun_soln_bbelow}
	y(z)=z^{2\tau}+\alpha_1z^{2\tau+1}+\alpha_2z^{2\tau+2}+\alpha_3z^{2\tau+3}+\dots,
	\end{equation}
	where the $\alpha$'s are constants properly chosen so that the above is an eigenfunction of the Heun class operator. Similarly the bounded above representation hosts solution of the form
	\begin{equation}\label{heun_soln_babove}
	y(z)=z^{2\sigma}+\beta_1z^{2\sigma-1}+\beta_2z^{2\sigma-2}+\beta_3z^{2\sigma-3}+\dots
	\end{equation}
	where the $\beta$'s are chosen in the same manner as the $\alpha$'s.
	Application of the operator $J^0$ (Eq.~(\ref{newj0})) on any state $|j^0\rangle$ ($=kz^p$, where $k$ is the normalization constant) yields
	\begin{eqnarray}
	\nonumber J^0|j^0\rangle&=&\left[z\dfrac{d}{dz}-(\sigma+\tau)\right]kz^p\\
	\nonumber \implies j^0|j^0\rangle&=&(p-\sigma-\tau)kz^p=(p-\sigma-\tau)|j^0\rangle\\
	\implies j^0&=&p-\sigma-\tau. \label{weight_power_reln}
	\end{eqnarray}
	Hence for the highest weight $|j^0_{max}\rangle$ ($=k_{max}z^{p_{max}}$ where $k_{max}$ is the normalization constant and $p_{max}(=2\sigma)$ is the highest monomial power)
	\begin{eqnarray}
	j^0_{max}&=&\sigma-\tau=j \ (\mbox{say}).
	\end{eqnarray}
	Similarly application of $J^0$ on the lowest weight state $|j^0_{min}\rangle$ gives
	\begin{equation}
	j^0_{min}=-(\sigma-\tau)=-j.
	\end{equation}
	Thus, for real values of $j$, the bounded above and the bounded below solution spaces are symmetrically placed about $j=0$, along the Casimir horizontal line corresponding to the constant given by Eq.~(\ref{construct_casimir}. The states of these representations are monomials with the powers differing from the highest (lowest) state by natural numbers (the weights differ by natural numbers according to the ladder operator structure and Eq.~(\ref{weight_power_reln}) implies that the corresponding monomial powers too change by the same natural numbers). The parameter $j$ will be called the \emph{representation parameter} since it is the deciding factor as to which representation spaces are available as the solution spaces for a particular equation, as to be discussed now.
	It is clear from Fig.~\ref{fig_SU11} that for $(\sigma-\tau)=j < 0$, the bounded below and bounded above representations will be of PD and ND types respectively and two power series solutions of the forms~(\ref{heun_soln_bbelow}) and (\ref{heun_soln_babove}) respectively, will be obtained from them. For $(\sigma-\tau)=j \geq 0$, the bounded below and above representations will be of NBB and NBA types respectively, and once again will host power series solutions of the forms~(\ref{heun_soln_bbelow}) and (\ref{heun_soln_babove}) respectively. It is also clear that for the latter case, there will be overlap between the Casimir lines corresponding to the two representations. A special case of this occurs when $j$ is a non-negative half integer, i.e., $j=N/2$ (where $N \in \mathbb{N}_0$). In that case $j$ and $-j$ differ by integer $N$ implying that there is a $(N+1)$ states overlap between the two representations and the two end states of these $(N+1)$ states lie on the $g(j^0)$ and the $g(j^0-1)$ curves. Or, in more simple terms, both the infinite dimensional bounded above and the bounded below representations split up into two pieces--- one of which is finite $(N+1)$ dimensional and another is infinite dimensional, the finite piece being common to both. The finite dimensional representation hosts $(N+1)$ independent solutions of the form
	\begin{equation}\label{heun_soln_finite_dim}
	y(z)=z^{2\tau}+\alpha_1z^{2\tau+1}+\dots+\alpha_{N-1}z^{2\tau+N-1}+\alpha_Nz^{2\sigma}=z^{2\tau}P_N(z),
	\end{equation}
	for $(N+1)$ eigenvalues. The eigenvalues and the $\alpha$'s are found by solving the eigenvalue problem of the $(N+1)$ dimensional matrix form of the Heun operator in the basis formed by the $(N+1)$ monomials spanning the finite dimensional representation. These types of solutions are the (quasi-)polynomials, which are of central interest in this paper. For other values of the eigen-parameter., one can find two power series solutions from the two infinite dimensional reducible representations consisting of the two irreducible pieces described above.
	
	For complex values of $j$, the representations can not be visualized by figures like Fig.~\ref{fig_SU11}. However, the fact that $z^{2\tau}$ and  $z^{2\sigma}$ are killed by $J^-$ and $J^+$ respectively, ensures that infinite dimensional bounded below and bounded above representations are available as the Heun solution spaces, which host power series solutions of the forms~(\ref{heun_soln_bbelow}) and (\ref{heun_soln_babove}) respectively.
	
	\newpage
	\section{Connection with the Frobenius solutions and with the solutions using the generators~(\ref{jpjzjm})}\label{appendix_imp_connections}
	
	The connection between the solutions to the Heun class of equations obtained using their bi- parametric $su(1,1)$ structure presented in this paper and the standard Frobenius type solutions is an interesting aspect of study. It may also seem relevant to ask whether there exists any map between the additional solutions found using the present construct and the earlier set of generators~(\ref{jpjzjm}). This appendix deals with these two issues.
	
	\subsection{Connection with the Frobenius solutions}
	The algebraic solutions (see Appendix~\ref{appendix_rep_theory}) found to the Heun class of equations are either of the form
	\begin{equation}\label{heun_soln_bb}
	y(z)=z^{2\tau}+\alpha_1z^{2\tau+1}+\alpha_2z^{2\tau+2}+\alpha_3z^{2\tau+3}+\dots
	\end{equation}
	(where the $\alpha$'s are constants) or of the form
	\begin{equation}\label{heun_soln_ba}
	y(z)=z^{2\sigma}+\beta_1z^{2\sigma-1}+\beta_2z^{2\sigma-2}+\beta_3z^{2\sigma-3}+\dots
	\end{equation}
	where the $\beta$'s are constants or in the special case when $(\sigma-\tau)=N/2 \ (N \in \mathbb{N}_0)$, is of the form
	\begin{equation}\label{heun_soln_finite}
	y(z)=z^{2\tau}+\alpha_1z^{2\tau+1}+\dots+\alpha_{N-1}z^{2\tau+N-1}+\alpha_Nz^{2\sigma}=z^{2\tau}P_N(z)
	\end{equation}
	(where the $\alpha$'s are found by solving the $(N+1)$ dimensional matrix eigenvalue problem).
	
	A glance at the solutions given by Eqs.~(\ref{heun_soln_bb}) and (\ref{heun_soln_ba}) tells that these are of Frobenius power series type local solutions about the singularity $z=0$ and $z=\infty$ respectively. The solution given by Eq.~(\ref{heun_soln_finite}) is of truncated power series type which is local about both $z=0$ and $z=\infty$. Now let's check the exponents at $z=0$ and $z=\infty$ for the Heun class of equations. For this, the Heun class of equations given by Eqs.~\ref{heun_as_poly}) and (\ref{the_P's}) are written as
	\begin{equation}\label{lin_scnd_ordr_diff_eqn}
	\dfrac{d^2y}{dz^2}+p(z)\dfrac{dy}{dz}+q(z)y=0,
	\end{equation}
	where
	\begin{equation}
	p(z)=\dfrac{P_2(z)}{P_3(z)}=\dfrac{a_4z^2+a_5z+a_6}{a_0z^3+a_1z^2+a_2z+a_3} \quad \mbox{and} \quad q(z)=\dfrac{P_1(z)}{P_3(z)}=\dfrac{a_7z+a_8+a_9/z}{a_0z^3+a_1z^2+a_2z+a_3}.
	\end{equation}
	The indicial equation for the exponents at $z=0$, when it is a regular singular point, are given by~\cite{ronveauxbook,morse_Feshback}
	\begin{equation}\label{zero_indicial}
	\rho_0^2+(A_0-1)\rho_0+B_0=0,
	\end{equation}
	where $A_0=\lim_{z \to 0} zp(z)$ and $B_0=\lim_{z \to 0} z^2q(z)$. Now for the equations with regular singularity at $z=0$, one has $a_2 \neq 0$ and $a_3=0$ (section~\ref{subsec_algebraizability}), which gives $A_0=a_6/a_2$ and $B_0=a_9/a_2$ and thus 
	\begin{equation}\label{zero_reg_exp}
	\rho_0=\dfrac{a_2-a_6 \pm \sqrt{(a_2-a_6)^2-4a_2a_9}}{2 a_2}.
	\end{equation}
	Notice from Eq.~(\ref{tau_regular}) that for equations with regular singularity at $z=0$
	\begin{equation}
	\rho_0=2\tau.
	\end{equation}
	
	When $z=0$ is an irregular singular point, a linear indicial equation is found to exist if the equation is written in the canonical form. This means that of the two linearly independent local solutions about $z=0$, one will be of the power series (infinite or truncated) type. The indicial equation in this case takes the form\cite{morse_Feshback}
	\begin{equation}\label{zero_confluent_indicial}
	k_0\rho_0+l_0=0
	\end{equation}
	where $k_0$ and $l_0$ are the coefficients of the lowest powers of $z$ in $p(z)$ and $q(z)$ respectively. To find $k_0$ and $l_0$ it may first be noted that when $z=0$ is an irregular singular point, one has $a_2=a_3=0$ (section~\ref{subsec_algebraizability}). Thus, $p(z)$ and $q(z)$ becomes
	\begin{equation}
	p(z)=\dfrac{a_4z^2+a_5z+a_6}{z^2(a_0z+a_1)} \quad \mbox{and} \quad q(z)=\dfrac{a_7z^2+a_8z+a_9}{z^3(a_0z+a_1)}
	\end{equation}
	Hence, applying partial fraction one obtains $k_0=a_6/a_1$ and $l_0=a_9/a_1$. Thus the indicial equation.~(\ref{zero_confluent_indicial}) gives
	\begin{equation}\label{zero_irreg_exp}
	\rho_0=-a_9/a_6,
	\end{equation}
	which yields $\rho_0=2\tau$ for equation with irregular singularity at $z=0$ as well (Eq.~(\ref{the_thing_dhe})).
	
	A special case occurs when the whole $O^-$ part is absent, i.e., in addition to $a_2=a_3=0$ one also has $a_6=a_9=0$. In that case $p(z)$ and $q(z)$ becomes
	\begin{equation}
	p(z)=\dfrac{a_4z+a_5}{z(a_0z+a_1)} \quad \mbox{and} \quad q(z)=\dfrac{a_7z+a_8}{z^2(a_0z+a_1)}.
	\end{equation}
	Partial fraction of $p(z)$ and $q(z)$ shows that the lowest power of $z$ in them is $-1$ and $-2$ respectively, which is a property of regular singularity~\cite{morse_Feshback}. This means that the singularity at $z=0$ is regular when the $O^-$ part is absent. So the indicial equation will be quadratic and of the form~(\ref{zero_indicial}), where $A_0=a_5/a_1$ and $B_0=a_8/a_1$ and thus
	\begin{equation}\label{free_zero_exp}
	\rho_0=\dfrac{\left(a_1-a_5\right)\pm \sqrt{\left(a_1-a_5\right)^2-4a_1a_8}}{2a_1}.
	\end{equation}
	Now, $-a_8$ is the eigenvalue which can be chosen suitably for solving the equation, which implies the exponent $(\rho_0)$ of $z=0$ itself can be chosen suitably. Or, in other words, the exponent of $z=0$ is a function of the eigen-parameter. in this case. In section~\ref{subsec_algebraizability}, the parameter $\tau$ has been found to be a free parameter for the equations with the $O^-$ part absent. So one can say that $\rho_0=2\tau$ holds for this case as well.
	
	To find the exponents at $z=\infty$, let's first effect the transformation $z \rightarrow 1/t$ to the Heun class general equation.~(\ref{lin_scnd_ordr_diff_eqn}) to obtain
	\begin{equation}
	\dfrac{d^2y}{dt^2}+P(t)\dfrac{dy}{dt}+Q(t)y=0
	\end{equation}
	where,
	\begin{subequations}
		\begin{align}
		P(t)=&\dfrac{2}{t}-\dfrac{1}{t^2}p\left(\dfrac{1}{t}\right)=\dfrac{2}{t}-\dfrac{a_4+a_5t+a_6t^2}{t(a_0+a_1t+a_2t^2+a_3t^3)} \\ \mbox{and} \quad Q(t)=&\dfrac{1}{t^4}q\left(\dfrac{1}{t}\right) \qquad =\dfrac{a_7+a_8t+a_9t^2}{t^2(a_0+a_1t+a_2t^2+a_3t^3)}.
		\end{align}
	\end{subequations}
	For regular singularity at $z=\infty$, the indicial equation takes form similar to Eq.~(\ref{zero_indicial}) with $A_{\infty}=2-a_4/a_0$ and $B_{\infty}=a_7/a_0$, which gives
	\begin{equation}\label{inf_reg_exp}
	\rho_{\infty}=\dfrac{(a_4-a_0)\pm\sqrt{(a_4-a_0)^2-4a_0a_7}}{2a_0}.
	\end{equation}
	A comparison with Eq.~(\ref{sigma&cp}) gives
	\begin{equation}
	\rho_{\infty}=-2\sigma
	\end{equation}
	for equation with regular singularity at $z=\infty$. When the singularity at $z=\infty$ is irregular, a linear indicial equation of the form of Eq.~(\ref{zero_confluent_indicial}) is found to exist. To find $k_{\infty}$ and $l_{\infty}$ one first notes that $a_0=0$ for all equations with irregular singularity at $z=\infty$ (section~\ref{subsec_algebraizability}). Application of partial fraction gives lowest power of $t$ in $P(t)$ as $k_{\infty}=a_4/a_1$ and the coefficient of the lowest power of $t$ in $Q(t)$ as $l_{\infty}=a_7/a_1$. This gives
	\begin{equation}\label{inf_irreg_exp}
	\rho_{\infty}=-\dfrac{a_7}{a_4},
	\end{equation}
	which means $\rho_{\infty}=-2\sigma$ holds (see Eq.~(\ref{sigma_cnflnt})) for all equations with irregular singularity at $z=\infty$ as well.
	
	A special case occurs when the whole $O^+$ part is absent, i.e., in addition to $a_1=0$ one also has $a_4=a_7=0$. This does not happen for the only equation (THE) with $a_3 \neq 0$, since for the THE: $a_4=-3 \neq 0$. So $a_3=0$ can be assumed for the cases where $O^+$ is absent. In that case $P(t)$ and $Q(t)$ becomes
	\begin{equation}
	P(t)=\dfrac{2}{t}-\dfrac{a_5+a_6t}{t(a_1+a_2t)} \quad \mbox{and} \quad Q(t)=\dfrac{a_8+a_9t}{t^2(a_1+a_2t)}.
	\end{equation}
	Partial fraction of $P(t)$ and $Q(t)$ shows that the lowest power of $t$ in them is $-1$ and $-2$ respectively, which is a property of regular singularity~\cite{morse_Feshback}. This means that the singularity at $z=\infty$ is regular when $O^+$ is absent. So the indicial equation will be quadratic and of the form~(\ref{zero_indicial}), with $A_{\infty}=2-a_5/a_1$ and $B_{\infty}=a_8/a_1$ and thus
	\begin{equation}\label{free_inf_exp} 
	\rho_{\infty}=\dfrac{\left(a_5-a_1\right)\pm \sqrt{\left(a_5-a_1\right)^2-4a_1a_8}}{2a_1}.
	\end{equation}
	Now, $-a_8$ is the eigenvalue which can be chosen suitably for solving the equation, which implies the exponent $(\rho_{\infty})$ of $z=\infty$ itself can be chosen suitably. Or, in other words, the exponent of $z=\infty$ is a function of the eigen-parameter. in this case. In section~\ref{subsec_algebraizability}, the parameter $\sigma$ has been found to be a free parameter for the equations with the $O^-$ part absent. So one can say that $\rho_{\infty}=-2\sigma$ holds for this case as well.
	
	Thus it is found that the parameters $\tau$ and $\sigma$ of the proposed algebraic construct for the Heun class of equations are practically half of the exponent value(s) at $z=0$ and negative of half of the exponent value(s) at $z=\infty$, respectively. This means that the algebraic bounded below (above) solution with lowest monomial power $2\tau$ (highest monomial power $2\sigma$) is actually the Frobenius solution about $z=0$ ($z=\infty$). This also justifies the fact that just a single value of $\tau$ or $\sigma$ is obtained for equations with irregular singularity at $z=0$ or $z=\infty$ respectively, because at most one local power series solution is possible for irregular singularities. One local power series solution about an irregular singularity is not guaranteed for differential equations in general. It is interesting to note that the Heun class equations in their canonical forms, always admit at least one local power series solution with non-essential singularity about both $z=0$ and $z=\infty$, even when the singularities at these two points are irregular (since one exponent is always obtained, as given by Eqs.~(\ref{zero_irreg_exp}) and (\ref{inf_irreg_exp}, for irregular singularities at $z=0$ and $z=\infty$, respectively). However, the real benefit of using algebraic structure is realized in case of the quasi-polynomials. When $(\sigma-\tau)$ is a non-negative half integer $N/2$ (which is the condition for obtaining quasi-polynomials), the lowest power $(2\tau)$ in the local Frobenius solution about $z=0$ is smaller than the highest power $(2\sigma)$ in the local Frobenius solution about $z=\infty$ by an integer $N$. In other words, the statement $(\sigma-\tau)= N/2$, $N \in \mathbb{N}_0$ is equivalent to saying that there are $(N+1)$ common monomials in the power series solutions about $z=0$ and $z=\infty$. This monomial overlap is the necessary condition for having truncated (finite number of terms) Frobenius solution(s) of the form $z^{2\tau}P_N(z)$ for any differential equation, however it is not a sufficient condition. The sufficiency comes from the algebraic construct which has shown that for Heun class of equations whenever there is this monomial overlap (or $(\sigma-\tau)= N/2$, $N \in \mathbb{N}_0$), a finite dimensional representation of $su(1,1)$ is available as an invariant subspace of the Heun operator in concern. The truncated power series (or quasi-polynomial) solutions are valid around both $z=0$ and $z=\infty$ and thus valid throughout the complex plane with proper branch cuts to avoid the other finite singular points. This is why a quasi-polynomial solution is of such utter importance.
	
	Thus the bounded one side infinite dimensional representations (either PD and ND type or NBB and NBA type,  depending on whether the value of the representation parameter is positive or negative) are found to host the standard Frobenius series solutions about $z=0$ and $z=\infty$, whereas, the bounded both side finite dimensional representations are found to host the truncated Frobenius series or (quasi-)polynomial type solutions, whenever they are admitted by the equation.
	
	\subsection{Connection with the solutions obtained using the earlier set of generators}\label{subsec_appB_Tur_connection}
	
	As discussed earlier, the generators.~(\ref{newgenerators}) with $0$ value of the parameter $\tau$ are actually the generators~(\ref{jpjzjm}). Hence, whenever for a Heun class equation one has $\tau=0$, the corresponding solutions can be obtained using the generators~(\ref{jpjzjm}). Now it has been seen earlier that with the use of the generators~(\ref{newgenerators}) there is possibility of finding non-zero $\tau$ (Eqs. (\ref{tau_regular}) and (\ref{the_thing_dhe}). Evidently, there is no direct way to find the corresponding solutions using the generators~(\ref{jpjzjm}). However, if one effects the F-homotopic transformation
	\begin{equation}\label{f_homotopy}
	y(z)=z^{2\tau}\xi(z)
	\end{equation}
	with the non-zero $\tau$ value to a Heun class equation, then the exponents $(0, \ 2\tau)$ at $z=0$ are shifted to $(-2\tau, \ 0)$ and the solutions corresponding to the previous exponent $2\tau$ can now be directly tracked with the generators~(\ref{jpjzjm}) but obviously the solutions corresponding to the previous exponent $0$ (now, $-2\tau$) is not now directly obtainable using the generators~(\ref{jpjzjm}). Using the generators~(\ref{newgenerators}), all the solutions of the power series or polynomial type can be obtained simultaneously.
	
	If there is only one value of $\tau$ which is not $0$, as may be the case with DHE (Eq.~(\ref{tau_dhe})), then the equation is directly algebraizable with generators~(\ref{newgenerators}) but not with the generators~(\ref{jpjzjm}). However, applying the transformation~(\ref{f_homotopy}) to the concerned DHE, makes it algebraizable using generators~(\ref{jpjzjm}) as well.

	\newpage
	
	\begin{landscape}
		\section{Matrix forms of GHE, CHE and BHE when they admit $(N+1)$ linearly independent polynomial solutions}\label{appendix_poly_conds}
		\begin{equation}\label{GHE_tridiag_poly}
		\mathcal{H}^p_{GHE}=
		\begin{bmatrix}[1.5]
		0 & \substack{a\gamma} & 0 & 0&\dots&0&0 \\
		
		\substack{\alpha\beta} & \substack{-(a(\delta+\gamma)+\epsilon+\gamma)} &  \substack{2a(1+\gamma)}& 0&\dots&0&0 \\
		
		0     & \substack{\alpha\beta+(\gamma+\epsilon+\delta)}     & \substack{-2(a+1)-2(a(\delta+\gamma)+\epsilon+\gamma))}&\substack{3(-2(a(\delta+\gamma)+\epsilon+\gamma)+a\gamma)}&\dots&0&0\\
		
		\vdots&\vdots&\vdots&\vdots&\vdots&\vdots&\vdots\\ 
		
		0&0&0&0&\dots&\substack{\alpha\beta+(N-1)[(N-2)\\-(\gamma+\epsilon+\delta)]}&\substack{-N(N-1)(1+a)\\-N(a(\delta+\gamma)+\epsilon+\delta)}
		\end{bmatrix}
		\end{equation}
		
		\vspace{.5cm}
		
		\begin{equation}\label{CHE_tridiag_poly}
		\mathcal{H}^{p}_{CHE}=
		\begin{bmatrix}[1.8]
		0
		&\substack{1-\kappa}&0&0&\dots&0&0\\
		
		\substack{-N\kappa}&\substack{(\gamma+\delta-\kappa)}&\substack{2(1-\kappa)}&0&\dots&0&0\\
		
		0&\substack{-(N-1)\kappa}&\substack{2(1+\gamma+\delta-\kappa)}&\substack{3(1-\kappa)}&\dots&0&0\\
		
		\vdots&\vdots&\vdots&\vdots&\vdots&\vdots&\vdots\\
		
		0&0&0&0&\dots&\substack{-\kappa}&\substack{N(N-1+\gamma+\delta-\kappa)}
		\end{bmatrix}
		\end{equation}
		
		\vspace{.5cm}
		
		\begin{equation}\label{BHE_tridiag_poly}
		\mathcal{H}^{p}_{BHE}=
		\begin{bmatrix}[1.8]
		0 & \substack{(\alpha+1)} & 0 & 0&\dots&0& 0&0 \\
		\substack{(\gamma-\alpha-2)} & \substack{-\beta} &  \substack{2(\alpha+2)}& 0&\dots&0&0&0 \\
		0     & \substack{(\gamma-\alpha-2)-2}  & \substack{-2\beta}&\substack{3(\alpha+3)}&\dots&0&0&0\\
		\vdots&\vdots&\vdots&\vdots&\vdots&\vdots&\vdots&\vdots\\
		0&0&0&0&\dots&\substack{(\gamma-\alpha-2)\\-2(N-2)}&\substack{(N-1)(N-2)-(N-1)\beta}&\substack{N(\alpha+N)}\\
		0&0&0&0&\dots&0&\substack{(\gamma-\alpha-2)\\-2(N-1)}&\substack{N(N-1)-N\beta}
		\end{bmatrix}
		\end{equation}
	\end{landscape}

\end{document}